\documentclass[twocolumn]{aastex7}

\usepackage{amsmath}	% Advanced maths commands
\usepackage{amssymb}	% Extra maths symbols
\usepackage{enumitem}
\usepackage{verbatim}   % for comment command
\usepackage[caption=false]{subfig}
\usepackage{array}
\usepackage{makecell}   % Allow line breaks within a table's cell with \makecell{} command
\usepackage{wrapfig}    % Wrap text around figure
\usepackage{listings}
\usepackage{hyperref}
\usepackage{cprotect}   % \verb in caption

\begin{document}

% Remove text indicating that this is a draft 
% \makeatletter\let\frontmatter@title@above=\relax

\title{The Roman View of Strong Gravitational Lenses}

\author[0000-0002-0748-7312]{Bryce Wedig}
\affiliation{Department of Physics, Washington University, St. Louis, MO 63130, USA}
\email{}

\author[0000-0002-6939-9211]{Tansu Daylan}
\affiliation{Department of Physics, Washington University, St. Louis, MO 63130, USA}
\affiliation{McDonnell Center for the Space Sciences, Washington University, St. Louis, MO 63130, USA}
\email{}

\author[0000-0003-3195-5507]{Simon Birrer}
\affiliation{Department of Physics and Astronomy, Stony Brook University, Stony Brook, NY 11794, USA}
\email{}

\author[0000-0002-7939-2988]{Francis-Yan Cyr-Racine}
\affiliation{Department of Physics and Astronomy, University of New Mexico, Albuquerque, NM 87106, USA}
\email{}

\author[0000-0003-1476-1241]{Cora Dvorkin}
\affiliation{Department of Physics, Harvard University, Cambridge, MA 02138, USA}
\email{}

\author[0000-0003-2808-275X]{Douglas P. Finkbeiner}
\affiliation{Department of Physics, Harvard University, Cambridge, MA 02138, USA}
\affiliation{Harvard-Smithsonian Center for Astrophysics, 60 Garden St., Cambridge, MA 02138, USA}
\email{}

\author[0009-0001-8629-8826]{Alan Huang}
\affiliation{Department of Physics and Astronomy, Stony Brook University, Stony Brook, NY 11794, USA}
\email{}

\author[0000-0001-8156-0330]{Xiaosheng Huang}
\affiliation{Department of Physics and Astronomy, University of San Francisco, San Francisco, CA 94117, USA}
\email{}

\author[0009-0007-6289-4360]{Rahul Karthik}
\affiliation{Department of Physics and Astronomy, Stony Brook University, Stony Brook, NY 11794, USA}
\email{}

\author[0000-0001-5512-2716]{Narayan Khadka}
\affiliation{Department of Physics and Astronomy, Stony Brook University, Stony Brook, NY 11794, USA}
\email{}

\author[0000-0002-5554-8896]{Priyamvada Natarajan}
\affiliation{Department of Astronomy, Yale University, 219 Prospect Street, New Haven, CT 06511, USA}
\affiliation{Department of Physics, Yale University, 217 Prospect Street, New Haven, CT 06511, USA}
\email{}

\author[0000-0001-6809-2536]{Anna M. Nierenberg}
\affiliation{University of California, Merced, 5200 N Lake Road, Merced, CA 95341, USA}
\email{}

\author[0000-0002-8040-6785]{Annika H. G. Peter}
\affiliation{Department of Physics, The Ohio State University, 191 W. Woodruff Ave, Columbus, OH 43210, USA}
\affiliation{Department of Astronomy, The Ohio State University, 140 W. 18th Ave, Columbus, OH 43210, USA}
\affiliation{Center for Cosmology and AstroParticle Physics, The Ohio State University, 191 W. Woodruff Ave, Columbus, OH 43210, USA}
\email{}

\author[0000-0002-2361-7201]{Justin D. R. Pierel}
\affiliation{Space Telescope Science Institute, 3700 San Martin Drive, Baltimore, MD 21218, USA}
\affiliation{NASA Einstein Fellow}
\email{}

\author[0009-0007-3185-7030]{Xianzhe TZ Tang}
\affiliation{Department of Astronomy, Boston University, 685 Commonwealth Ave., Boston, MA 02215, USA}
\affiliation{Department of Physics and Astronomy, Stony Brook University, Stony Brook, NY 11794, USA}
\email{}

\author[0000-0003-2229-011X]{Risa H. Wechsler}
\affiliation{Kavli Institute for Particle Astrophysics and Cosmology and Department of Physics, Stanford University, Stanford, CA 94305, USA}
\affiliation{Department of Particle Physics \& Astrophysics, SLAC National Accelerator Laboratory, Menlo Park, CA 94025, USA}
\email{}

\begin{abstract}

Galaxy-galaxy strong gravitational lenses can constrain dark matter models and the Lambda Cold Dark Matter cosmological paradigm at sub-galactic scales. Currently, there is a dearth of images of these rare systems with high signal-to-noise and angular resolution. The \textit{Nancy Grace Roman Space Telescope} (hereafter, \textit{Roman}), scheduled for launch in late 2026, will play a transformative role in strong lensing science with its planned wide-field surveys. With its remarkable 0.281 square degree field of view and diffraction-limited angular resolution of $\sim$0.1 arcsec, \textit{Roman} is uniquely suited to characterizing dark matter substructure from a robust population of strong lenses. We present a yield simulation of detectable strong lenses in \textit{Roman's} planned High Latitude Wide Area Survey (HLWAS). We simulate a population of galaxy-galaxy strong lenses across cosmic time with Cold Dark Matter subhalo populations, select those detectable in the HLWAS, and generate simulated images accounting for realistic Wide Field Instrument detector effects. For a fiducial case of single 146-second exposures, we predict around 160,000 detectable strong lenses in the HLWAS, of which about 500 will have sufficient signal-to-noise to be amenable to detailed substructure characterization. We investigate the effect of the variation of the point-spread function across \textit{Roman's} field of view on detecting individual subhalos and the suppression of the subhalo mass function at low masses. Our simulation products are available to support strong lens science with \textit{Roman}, such as training neural networks and validating dark matter substructure analysis pipelines.

\end{abstract}

%-----------------------------------------------------------------------
\section{Introduction}
\label{section:Introduction}

Strong gravitational lensing by galaxies has broad applications in astrophysics and cosmology, such as mapping dark matter on sub-galactic scales \citep{McKean2007, Vegetti2010, Vegetti2024}, providing an independent probe of the Hubble constant $H_0$ through time-delay cosmography \citep{Suyu2017, Treu2024, Birrer2024}, measuring the mass profiles of galaxies to test galaxy evolution models \citep{Koopmans2006, Shajib2021}, and studying magnified high-redshift objects \citep{Marshall2007, Agol2009}. Dark matter substructure in individual galaxies and along the line-of-sight offer a test of the Lambda cold dark matter ($\Lambda$CDM) cosmological model on sub-galactic scales, as $\Lambda$CDM makes testable predictions for the properties of substructure that can be constrained by strong lensing, such as the subhalo mass function and subhalo concentration. Alternative dark matter models, such as warm dark matter \citep[WDM;][]{Bode2001} and self-interacting dark matter \citep{Spergel2000}, predict that substructure will deviate from $\Lambda$CDM in the mass function and halo density profiles at sub-galactic scales \citep{Colin2000, Rocha2013}.

\begin{figure}[htb!]
    \centering
    \plotone{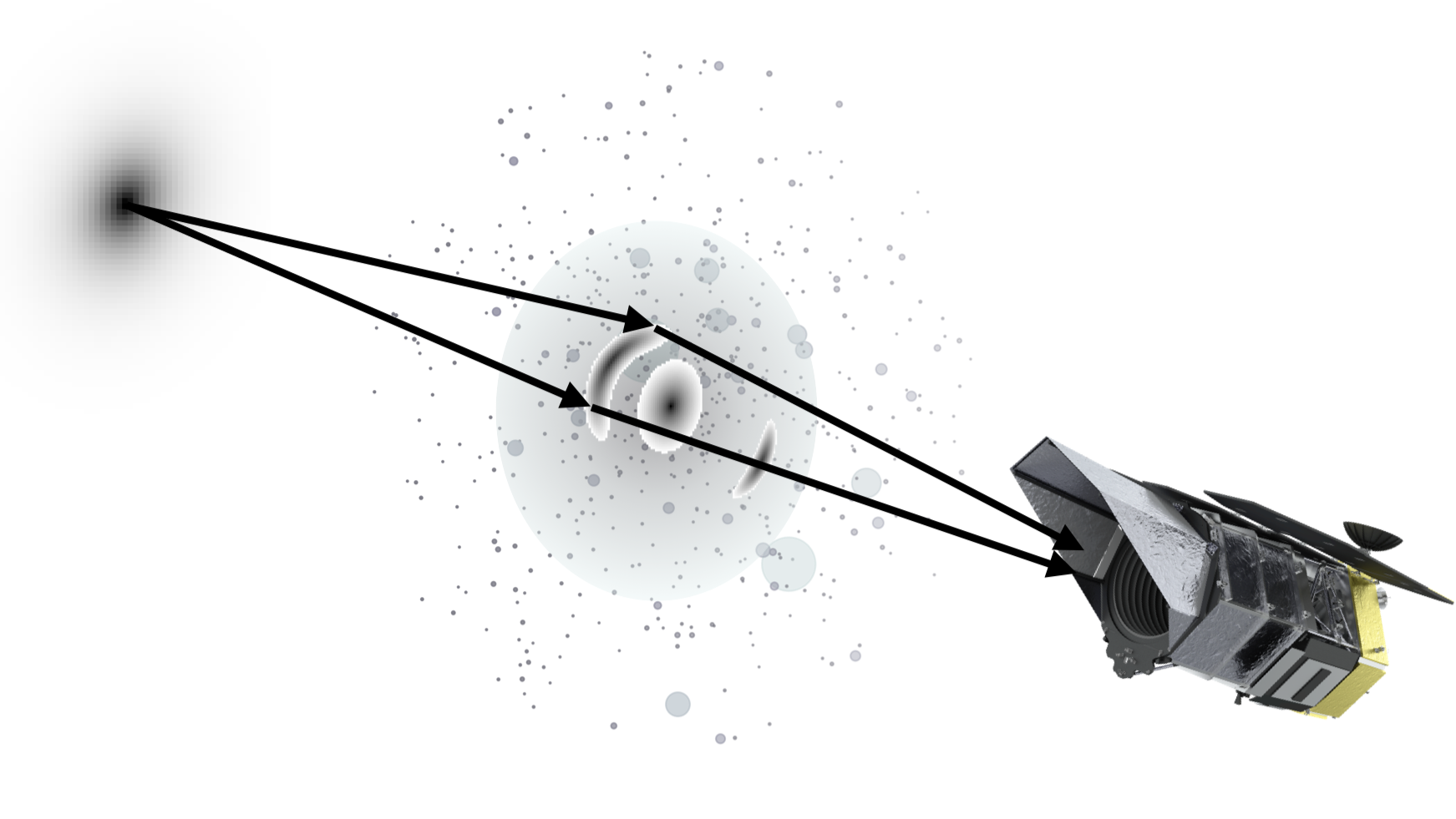}
    \caption{The galaxy-galaxy strong lensing configuration. Light from a distant source galaxy is gravitationally lensed by the foreground lensing galaxy, forming multiple distorted and magnified images of the source. The lensing potential is due to a main halo and a population of subhalos in the lens plane. Additionally, shear and external convergence are added to the mass model to simulate the effect of massive, distant line-of-sight subhalos.}
    \label{fig: cartoon}
\end{figure}

Strong lensing is uniquely well-suited to addressing this issue since subhalos below roughly $10^7$ to $10^8$ M$_\odot$ are not expected to have luminous counterparts \citep{Bullock2000, Somerville2002, Nadler2021, Ahvazi2024}. Thus, their only observable signatures would be through their gravitational effects. While other probes of dark matter substructure exist, including signatures in the power spectrum of the Lyman-$\alpha$ forest at small scales \citep[e.g.,][]{Irsic2017, Villasenor2023}, perturbations of stellar streams \citep[e.g.,][]{Erkal2017, Bonaca2019}, and observations of Milky Way satellite galaxies \citep[e.g.,][]{Kim2021, Nadler2021, Newton2021}, statistical analyses of images of strong lenses using forward-modeling methods \citep[e.g.,][]{Vegetti2012, Birrer2017, Daylan2018, He2022} and Bayesian convolutional neural networks \citep[e.g.,][]{WagnerCarena2024, Tsang2024} offer powerful approaches to constraining the properties of substructure in galaxies. The galaxy-galaxy strong lensing configuration we consider in this work is illustrated in Figure \ref{fig: cartoon}.

Currently, science with strong lenses is limited by the small number of known systems and the quality of their images. Roughly 1500 galaxy-galaxy strong lenses have been confirmed to date \citep{Lemon2024} and the number of candidates is approaching 10$^4$ \citep[e.g.,][]{Petrillo2019, Huang2020, Huang2021, Storfer2024}, but nearly all have only been imaged by ground-based instruments such as the Dark Energy Camera \citep{Flaugher2015} and VLT Survey Telescope \citep{Capaccioli2011} and are therefore seeing-limited. Sensitivity to the signature of subhalos depends critically on high angular resolution, signal-to-noise, and the separation of source galaxy arcs and lensing galaxy light, i.e., strong lenses with large Einstein radii ($\gtrsim1$ arcsec) \citep{Despali2022}. The effect of lower-mass subhalos is small: the angular deflection at the scale radius of a modestly-sized $10^8$ M$_\odot$ subhalo is expected to be of order milli-arcseconds. Additionally, degeneracies in the mass model between the smooth main halo and substructure \citep{He2023, Nightingale2024}, simplistic light models \citep{Nightingale2024}, and unmodeled baryonic structures in the lensing galaxy \citep{Gilman2017, Hsueh2018} contribute to uncertainties in substructure inference.

Space-based observations offer the angular resolution required to be sensitive to lower-mass subhalos. Notable sources of the current sample of space-based imagery of strong lenses include the \textit{Hubble Space Telescope} (\textit{HST}), and more recently, \textit{JWST} \citep{Gardner2006}. \textit{HST's} population includes the CASTLES project \citep{Munoz1998, Falco1999} of 47 systems, Sloan Lens ACS Survey \citep[SLACS;][]{Bolton2008} of 131 systems, Strong Lensing in the Legacy Survey (SL2S) project \citep{Sonnenfeld2013a} of 33 systems, BELLS GALLERY sample \citep{Shu2016a} of 21 systems, and around 100 systems in the \textit{HST} COSMOS field \citep{Koekemoer2007, Faure2008, Jackson2008, Pourrahmani2018}. \citet{Garvin2022} report 252 candidates from a search of all public \textit{HST} data through June 2020. \textit{JWST} observations of strong lenses include JWST-GO-2056 of 31 quadruply lensed quasars \citep{Nierenberg2024}, the TEMPLATES program of four systems \citep{Rigby2024}, and the serendipitously discovered and currently most distant Einstein ring JWST-ER1 \citep{vanDokkum2024}.

However, strong lens science is limited by the dearth of space-based observations of strong lenses. A single, massive subhalo that causes a high signal-to-noise perturbation of the lensed image can be detected through modeling the system, e.g., ``gravitational imaging'' \citep{Vegetti2010}. For single-subhalo detection in a high signal-to-noise system, the current detection threshold of \textit{HST} is estimated to be between $10^9$ and $10^{9.5}$ M$_\odot$ \citep{Amorisco2022, Despali2022}. There are only expected to be $\mathcal{O}(1)$ subhalos per typical lensing galaxy that are this massive. It is unlikely that such a subhalo will be sufficiently aligned with an image of the lensed source to be detectable via strong lensing. Since the number of available amenable strong lenses is small, only a few single subhalos have been detected in them (see Section \ref{subsection: subhaloDetectionCompare}) and most have given null results (e.g., no significant detections in a sample of 11 systems in \citet{Vegetti2014}, no significant detections in a sample of 17 systems in \citet{Ritondale2019}, and two significant detections in a sample of 54 systems in \citet{Nightingale2024}). The collective effect of many subhalos can also be studied using strong lenses, for instance, by measuring flux ratio anomalies \citep[e.g.,][]{Keeley2024, Nierenberg2024} or conducting population-level inference with machine learning \citep[e.g.,][]{WagnerCarena2024, Tsang2024}. Because inferring robust population-level statistics for subhalos requires many images of strong lenses, they have only been derived from simulated data \citep[e.g.,][]{Daylan2018, WagnerCarena2023} or a small sample (e.g., 23 systems in \citet{Fagin2024}). Strong lenses can place lower limits on the WDM thermal relic mass, e.g., using one system in \citet{Birrer2017}, eight systems in \citet{Gilman2020}, and 14 systems in \citet{Keeley2024}. The analysis in \citet{Birrer2017} takes a population-level, forward modeling approach, while \citet{Gilman2020} and \citet{Keeley2024} consider flux ratio anomalies. These constraints are comparable to those from other probes such as the Lyman-$\alpha$ forest \citep{Irsic2017, Irsic2024} and Milky Way satellite counts \citep{Newton2021, Nadler2021}, and are limited primarily by the statistics of the small number of amenable strong lenses \citep{Keeley2024}. Therefore, to place robust constraints on dark matter models, high angular resolution images of \emph{many} strong lenses are needed \citep{Vegetti2018, Vegetti2024}.

Upcoming and ongoing wide-field deep-sky surveys such as the Vera C. Rubin Observatory's Legacy Survey of Space and Time \citep[LSST;][]{Ivezic2019} and the \textit{Euclid Space Telescope's} Wide Survey \citep{Laureijs2011} are predicted to revolutionize strong lensing science by discovering $\mathcal{O}(10^5)$ new strong lenses \citep{Collett2015, Holloway2023, Ferrami2024}. Given the area and depth of these surveys, they are better suited for discovery. However, these instruments are not ideal for substructure characterization. Rubin is ground-based, so its angular resolution is seeing-limited ($\sim0.7$ arcsec), far above the threshold for detecting lower-mass subhalos. \textit{Euclid} is space-based so it has excellent angular resolution, but its visible instrument (VIS) has only one broad filter (0.55-0.9 microns). Color information is useful as it helps differentiate the light of the lensed source from the light of the lensing galaxy, and the signature of dark matter substructure is in the deflection of the lensed light. Since a galaxy's spectral energy distribution reddens over the course of its evolution due to the aging stellar population and dust, and spiral and irregular galaxies can have deceptive morphologies, wide spectral coverage aids in detecting and modeling strong lenses across cosmic time. \textit{Euclid's} Near Infrared Spectrometer and Photometer (NISP) instrument does have three filters, but its angular resolution is $\sim0.3$ arcsec.

The \textit{Nancy Grace Roman Space Telescope} \citep[\textit{Roman};][]{Spergel2015}, formerly known as the \textit{Wide-Field InfraRed Survey Telescope} (\textit{WFIRST}), is currently expected to launch in October 2026. \textit{Roman} will have two instruments: the Wide Field Instrument (WFI), equipped with eight imaging filters spanning 0.48-2.3 microns and two dispersive elements for spectroscopy, and the Coronagraph Instrument. As of January 2025, both instruments have been delivered to NASA's Goddard Space Flight Center for spacecraft integration and testing. 

\begin{figure}[htb!]
    \centering
    \includegraphics[width=0.35\textwidth]{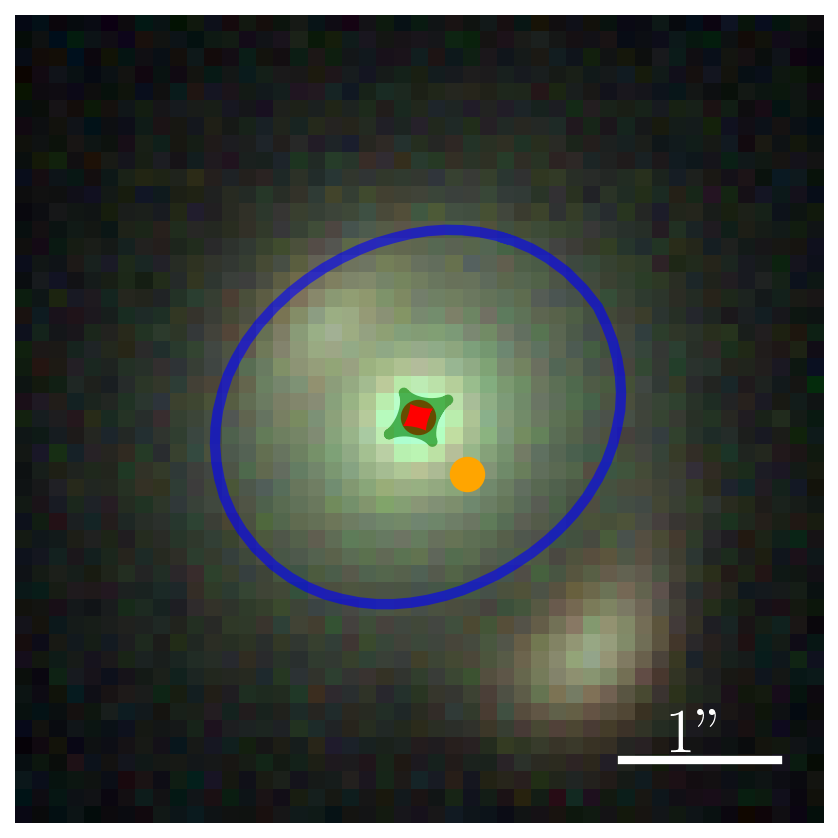}
    \caption{Simulated exposure of a sample strong lens detectable by \textit{Roman}: F184, F129, and F106 are mapped to red, green, and blue, respectively. The source and lens positions are denoted by the orange and red dots, respectively. The caustic and critical curves are denoted by the green and blue lines, respectively.}
    \label{fig: romanImage}
\end{figure}

\begin{figure}[htb!]
    \centering
    \plotone{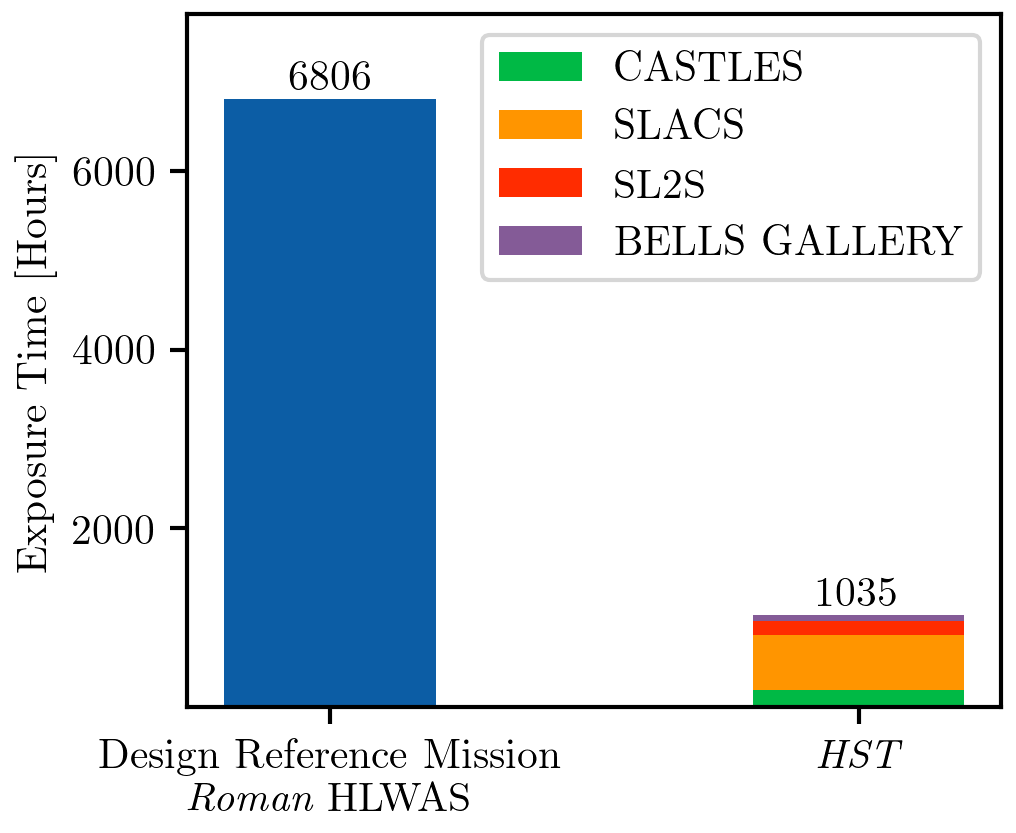}
    \caption{A rough comparison between \textit{Roman} and \textit{HST} of time spent observing strong lenses. We compare the planned duration of the HLWAS (left bar) with the total amount of time on \textit{HST} devoted to observing strong lenses as part of the CASTLES project \citep{Munoz1998, Falco1999}, Sloan Lens ACS Survey \citep{Bolton2008}, Strong Lensing in the Legacy Survey (SL2S) project \citep{Sonnenfeld2013a}, and BELLS GALLERY Survey \citep{Shu2016a} (right). The HLWAS design reference mission allocates 283.6 days to the imaging component. Since roughly 15 detectable strong lenses are expected in every exposure, the entire duration of the imaging component of the HLWAS can be considered time spent observing strong lenses.}
    \label{fig: exposureTimeComparison}
\end{figure}

\textit{Roman's} WFI imaging is well-suited to characterizing dark matter substructure from strong lenses due to its excellent angular resolution of $\sim$0.1 arcsec (see Figure \ref{fig: romanImage}). These filters extend to the near-infrared (NIR), well beyond \textit{Euclid's} VIS instrument, allowing \textit{Roman} to be sensitive to lensed images that dust in lensing galaxies could attenuate. \textit{Roman's} sensitivity and angular resolution are comparable to that of \textit{HST}. However, there is a tremendous gain, as \textit{Roman's} field of view is nearly 218 times larger than that of the IR channel of \textit{HST's} WFC3 instrument. As a result, despite their rarity, we predict around 27 detectable strong lenses in every exposure\footnote{We assume a single 146 s exposure.} of the design reference mission High Latitude Wide Area Survey \citep[HLWAS;][]{Dore2018}, with other forecasts varying between 1.3 and 14.6 per exposure (see Section \ref{subsection: SurveySimulationResults}). Figure \ref{fig: exposureTimeComparison} demonstrates how \textit{Roman's} wide field of view will enable the serendipitous collection of useful imaging in the HLWAS, improving dramatically on the data volume from strong lens surveys that \textit{HST} has been able to complete.

\textit{Roman's} strong lenses will enable significant advances in many strong lensing science cases. They will provide tighter constraints on the Hubble parameter and other cosmological parameters by discovering around 31 lensed supernovae \citep{Pierel2021}. Additionally, we predict that \textit{Roman's} sample of lensing galaxies will extend to higher redshifts than those from \textit{HST} (see Section \ref{subsection: SurveySimulationResults}). Measurements of the mass profiles of lensing galaxies to study galaxy evolution have been limited to $z<1$ \citep[e.g.,][]{Derkenne2021, Tan2024}, but we predict over 12,000 with $z\geq 1$ in the HLWAS.

In this work, we forecast \textit{Roman's} impact on studying dark matter substructure with galaxy-galaxy strong gravitational lenses with the purpose-built software pipeline \textsc{mejiro} \citep{Wedig2025}.\footnote{\url{https://github.com/AstroMusers/mejiro}, version 1.0.0} Doing strong lens science with \textit{Roman} will present unique challenges compared to previous instruments such as \textit{HST}, chief among which will be accounting for the variation of the point-spread function across the wide field of view. We aim to aid in the design and analysis of \textit{Roman} observations that will maximize its potential. We produce a dataset of simulated HLWAS exposures of strong lenses to be used as training and validation data for substructure inference pipelines.

The rest of this paper is structured as follows. Section \ref{section: Methodology} describes the simulation pipeline, from realistic galaxies to cutouts. In Section \ref{section: Results}, we discuss the results of our HLWAS strong lens yield simulation and the dataset of simulated images. In Section \ref{section: Discussion}, we consider the effect of \textit{Roman} systematics on discriminating between WDM and CDM and detecting individual subhalos. We compare \textit{Roman's} view of single subhalos to \textit{HST's}. We discuss planned improvements to this pipeline which will provide robust training data for neural networks and validate analysis pipelines for studying dark matter substructure with \textit{Roman} images of strong lenses. Section \ref{section: Conclusion} presents our conclusions.

Cosmological parameters from \citet{PlanckCollaboration2020} are adopted throughout this work.

%-----------------------------------------------------------------------
\section{Methodology}
\label{section: Methodology}

\subsection{Survey Simulation}
\label{subsection: SurveySimulation}

\subsubsection{Simulated Galaxy Population}
\label{subsubsection: SimulatedGalaxyPopulation}

To predict the population of strong lenses observable in \textit{Roman's} HLWAS, we start by simulating a population of early- and late-type galaxies out to redshift $z=6$. The light of the lens and source galaxies are assumed to be S\'ersic ellipse profiles. The morphologies are determined by \textsc{SkyPy}'s log-normal size distributions for early-type and late-type galaxies. Luminosities and spectra are determined by \textsc{SkyPy}'s Schechter luminosity function and Dirichlet-distributed spectral energy distribution modules. The parameters of these distributions are determined by the \textsc{SkyPy} and \textsc{SLSim} defaults for early- and late-type galaxies, respectively \citep{Shen2003, Bernardi2010}, or set in the \textsc{SkyPy} configuration file available in the \textsc{mejiro} repository.

The mass of the lensing galaxy is modeled with an elliptical power law (EPL) mass profile with shear and external convergence components. The shear and external convergence approximate the effect of line-of-sight structures and the distortions they produce, and their parameters are determined by \textsc{SLSim} default distributions (Tang et al., in prep.). The main halo mass is calculated following the empirical relationship between stellar mass and total lensing mass from \citet{Lagattuta2010}: 
\begin{equation}
    \frac{M_{\mathrm{vir}}(z)}{M_{*}(z)} = (51 \pm 36)(1 + z)^{(0.9 \pm 1.8)}.
\end{equation}
This redshift-dependent expression is obtained from a fit to a set of 41 strong lenses with lensing galaxies $0.4<z<0.9$ and 22 from \citet{Gavazzi2007} on $0.063<z<0.513$ from the SLACS set \citep{Bolton2008}. While this relationship is not derived from systems with lensing galaxies with $z>1$ where $<$8\% of \textit{Roman's} lensing galaxies are expected to be, in general, the stellar-to-halo-mass relation is not well-constrained at higher redshifts. Studies of the stellar-to-halo-mass relation using other probes suggest that it does not exhibit significant redshift-dependence below $z\sim1.5$ \citep[e.g.,][]{Wechsler2018, Shuntov2022, Zaidi2024}. Inaccuracies in this conversion would skew the subhalo populations, primarily in terms of subhalo abundance. The strong lenses best suited to substructure characterization that we use to derive the main results of this work (Section \ref{section: Discussion}) have lensing galaxies $z \lesssim 0.6$. We plan to improve the accuracy of high-redshift main halo masses in a future version of the pipeline by using a suite of hydrodynamic simulations to propagate the cosmological uncertainty in the mass-to-light ratio at high redshifts.

\subsubsection{Simulated Subhalo Populations}
\label{subsubsection: SimulatedSubhaloPopulations}

We expect subhalos to exist within the main halo due to hierarchical merging and accretion of dark matter predicted in $\Lambda$CDM. However, this process is substantiated by cosmological N-body simulations in a broader range of dark matter theories \citep{Moore1999, Bullock2017}. We assume CDM given its empirical success \citep[e.g.,][]{Hinshaw2013, Alam2017}. To generate realistic populations of CDM subhalos in the lensing galaxies of the detectable strong lenses, we use \textsc{pyHalo}\footnote{\url{https://github.com/dangilman/pyHalo}} \citep{Gilman2020}. \textsc{pyHalo} supports the generation of subhalo populations for a wide range of alternative dark matter theories. We utilize \textsc{pyHalo}'s preset \texttt{CDM} class \citep{Gilman2020, Gilman2022, Dhanasingham2023} to generate a population of CDM subhalos, where each subhalo is parameterized by a truncated NFW (TNFW) profile \citep{Baltz2009} given by the radial density profile
\begin{equation} \label{eq: tnfw}
    \rho(r) = \frac{r^2_{t}}{r^2 + r^2_{t}}\frac{\rho_0}{\frac{r}{r_s}\left(1 + \frac{r}{r_s}\right)^2}
\end{equation}
where $\rho_0$ is the density normalization, $r_t$ is the truncation radius, and $r_s$ is the scale radius. The subhalos are spatially distributed according to a cored NFW profile with a core radius that is half the scale radius of the main halo, matching the results of N-body cosmological simulations where tidal effects within this core radius result in a uniform number density of subhalos \citep{Gilman2020}. Following \citet{WagnerCarena2023}, for computational efficiency, we limit the generation of subhalos to within a disk of radius three times the Einstein radius around the center of the lensing galaxy. While N-body simulations suggest that subhalos exist beyond this range \citep{Zavala2019}, only those subhalos near or within the Einstein radius in projection will have any effect on the lensed emission. The subhalo masses ($M_{200}$ with respect to the critical density at $z=0$) range from $10^6$ to $10^{10}$ M$_\odot$ and are distributed via a subhalo mass function with logarithmic slope $\alpha=-1.9$ and normalization $\Sigma_{\textrm{sub}}=0.055\,\textrm{kpc}^{-2}$ \citep{Gilman2020}. The concentrations are sampled from the redshift-dependent mass-concentration relation of \citet{Diemer2019} implemented in \textsc{pyHalo}. \textsc{pyHalo} uses the results of a semi-analytic modeling code \textsc{Galacticus} \citep{Benson2012, Pullen2014} to determine how the subhalo mass function depends on parameters of the host such as main halo mass and redshift, rather than absorbing these dependencies into the overall normalization. This is especially important for this survey simulation, which includes lensing galaxies in the wide redshift range $0.01\leq z \leq3$. 

When adding subhalos to the mass model, the mass in subhalos is accounted for with a negative mass sheet at the redshift of the lensing galaxy. 

Halos along the line-of-sight are expected to introduce a non-trivial contribution to the lensing signature \citep{Despali2018, He2022, Dhanasingham2023, Hogg2023}. However, we do not simulate them here and leave it to future work (see Section \ref{subsection: FutureWork}).

\subsubsection{Roman's High Latitude Wide Area Survey} 
\label{subsubsection: HLWAS}

\textit{Roman} is expected to conduct three Core Community Surveys: the aforementioned HLWAS, the High Latitude Time Domain Survey (HLTDS), and the Galactic Bulge Time Domain Survey. While the general design of these surveys has been set to meet \textit{Roman's} science requirements, the details, including precise location, filters, exposure times, etc., will be defined by the community and are not yet set in stone. The HLWAS is relevant for imaging strong lenses as a wide ($\sim$1700 deg$^2$) survey of the extragalactic sky. It will utilize the WFI and is expected to have a spectroscopic component in addition to an imaging component. The imaging component is expected to utilize the F106, F129, F158, and F184 bands with an exposure time of $\sim$146 s and three dither positions per filter except for four positions for F129. We choose parameters for the simulated survey based on this possible design for the HLWAS.\footnote{\url{https://vmromanweb1.ipac.caltech.edu/page/high-latitude-wide-area-survey.html}}

The HLTDS, focused on supernova cosmology, is expected to conduct deep observations of the extragalactic sky over an area roughly one hundred times smaller than the HLWAS. We briefly discuss strong lenses in the HLTDS in Section \ref{subsection: LowestDetectableSubhaloMass}.

\subsubsection{Simulating the HLWAS}
\label{subsubsection: SimulatingDetectableLenses}

We simulate 171 deg$^2$ of the sky, 10\% of the planned 1700 deg$^2$ HLWAS. We utilize the strong lensing simulation pipeline \textsc{SLSim}\footnote{\url{https://github.com/LSST-strong-lensing/slsim}, version 0.1.0 commit \href{https://github.com/LSST-strong-lensing/slsim/tree/7ed0f7b53d4f23f583475e16c54a1a5a7edb68d6}{\texttt{7ed0f7b}}} (DESC and SLSC Collaboration, in prep.), which builds on the astronomical survey simulation package \textsc{SkyPy}\footnote{\url{https://github.com/skypyproject/skypy}} \citep{Amara2021, SkyPyCollaboration2023} to generate the parameters of realistic galaxies, then determines which will form strong lenses and calculates their properties. Source galaxies have redshift $0.01 \leq z \leq 6$, while lensing galaxies have redshift $0.01 \leq z \leq 3$. We filter out galaxies dimmer than 27th magnitude in \textit{Roman's} F129 filter (J band). We choose the F129 filter to allow for comparison with other \textit{Roman} strong lens survey forecasts in the literature (see Section \ref{subsection: SurveySimulationResults}). The cuts we place on this population of galaxies are summarized in Table \ref{table: surveyParams}. These cuts make the pipeline more efficient by removing any galaxies that have no chance of forming a detectable system prior to more computationally intensive constraints, such as estimating the signal-to-noise ratio (SNR) from a simulated exposure. 

\textit{Roman} filter responses are calculated using filter effective area curves dated March 2024\footnote{\url{https://roman.gsfc.nasa.gov/science/WFI_technical.html}, version 1\label{footnote: RomanTechnicalDocumentation}} and the spectroscopic data utility package \textsc{speclite}\footnote{\url{https://github.com/desihub/speclite}} \citep{Kirkby2024}. Filter effective area curves are provided for each of the WFI's 18 detectors, as there are variations in bandpasses and quantum efficiency among them. These variations, coupled with differences in wavefront error across the large focal plane array (see Figure \ref{fig: psfsAcrossFocalPlane}), are important systematics to account for when doing substructure characterization as we demonstrate in Section \ref{section: Discussion}. We use detector-specific zero-point magnitudes provided in the \texttt{roman-technical-information} GitHub repository\footnote{\url{https://github.com/spacetelescope/roman-technical-information}, version 1.0 commit \href{https://github.com/spacetelescope/roman-technical-information/tree/16db38b22e234277cbb65c42f4dd6a58a2eaec96}{16db38b}} managed by the Space Telescope Science Institute (STScI). We incorporate detector-dependent filter responses into our simulation pipeline by observing an equal sky area with each detector using its effective area curve and performing flux calculations with the appropriate zero-point magnitude.

\begin{deluxetable}{ll}[htb!]
    \tablenum{1}
    \tablecaption{Parameters for the Simulated \textit{Roman} HLWAS\label{table: surveyParams}}
    \tablewidth{0pt}
    \tablehead{
        \multicolumn1l{Name} & \multicolumn1l{Value}}
    \startdata
    Total Simulated Survey Area & 171 deg$^2$ \\
    \textit{Roman} Filters & F106, F129, F158, F184 \\
    Exposure Time & 146 s \\
    Source Redshifts & $[0.01, 6]$ \\
    Lens Redshifts & $[0.01, 3]$ \\
    Source Magnitudes & $m_\textrm{F129}<27$ \\
    Lens Magnitudes & $m_\textrm{F129}<27$ \\
    \enddata
\end{deluxetable}

\subsubsection{Detectability Criteria}
\label{subsubsection: DetectabilityCriteria}

We identify pairs of source and lensing galaxies that form galaxy-galaxy strong gravitational lenses based on the following criteria:
\begin{enumerate}
    \item The redshift of the source galaxy is greater than the redshift of the lensing galaxy.
    \item The angular distance between source and lensing galaxies along our line-of-sight must be less than $\sqrt{2}$ times the Einstein radius of the configuration of those galaxies.
    \item More than one image is formed.
\end{enumerate}
While the source and lensing galaxies may be sufficiently aligned to create a strong lens, the images may be too dim or close together for the system to be detectable. For a strong lens to be \emph{detectable}, it must also meet the following criteria:
\begin{enumerate}
    \setcounter{enumi}{3}
    \item Twice the Einstein radius is greater than or equal to 0.3 arcsec.
    \item The distance between images for at least one pair of image positions is greater than or equal to 0.3 arcsec.
    \item The magnitude of the magnified source is brighter than 25th magnitude (F129).
    \item The unlensed flux from the source galaxy is magnified by more than a factor of three.
    \item The SNR of the lensed images is greater than 20 in F129.
\end{enumerate}
The first six conditions are built into \textsc{SLSim}'s \texttt{draw\_population} method that determines which galaxies form detectable strong lenses given a population of galaxies. The magnification criterion, a typical criterion of strong lens detectability, ensures that the images are sufficiently magnified. The SNR threshold criterion filters out systems that are not distinct enough to be identified by lens-finding algorithms due to the lensed images being washed out by the light of the lensing galaxy. Our definition of SNR accounts for contamination from the lensing galaxy by counting it as noise (see Equation \ref{eq:SNR}). $20$ is a typical value for this cutoff \citep[e.g.,][]{Collett2015, Weiner2020, Holloway2023, ORiordan2023}. The particular SNR threshold for a high-quality strong lens will depend on the science case. For the purposes of dark matter subhalo detection, \citet{Despali2022} shows that the lowest detectable subhalo mass decreases linearly with SNR.

To estimate SNR, we generate simulated exposures of each strong lens using the full image simulation pipeline described in Section \ref{subsection: ImageSimulation}. Again, we choose the F129 filter to allow for comparison with other \textit{Roman} strong lens forecasts. We use a point-spread function (PSF; see Section \ref{subsubsection: CalculatingObservedImage} for details of the PSF model) at the center of the detector and ray-shoot, assuming a smooth macromodel for computational efficiency. We first calculate the SNR in each pixel by dividing a simulated exposure of the lensed emission (i.e., the signal) by the square root of the total simulated exposure (i.e., the noise for a Poisson process). Following \citet{Holloway2023}, we define regions from groups of adjacent pixels where SNR $>$ 1, then calculate the SNR of each region according to
\begin{equation}\label{eq:SNR}
    \textrm{SNR}_\textrm{region} = \frac{\sum\limits_i N_{i,\,S}}{\sqrt{\sum\limits_i \left(N_{i,\,S} + N_{i,\,L} + N_{i,\,B} + N_{i,\,N}\right)}}
\end{equation}
where the summations are over the pixels that comprise the region, $N_{i,\,S}$ are the counts in pixel $i$ due to the source galaxy, $N_{i,\,L}$ are counts due to the lensing galaxy, and $N_{i,\,B}$ are counts due to sky background, and $N_{i,\,N}$ are counts due to detector noise. If multiple regions are formed, we take the SNR of the region with the highest SNR to be the SNR of the system. For the planned exposure time of the HLWAS, this method of SNR per pixel greater than unity reliably isolates the lensed emission across the range of strong lens morphologies.

\subsection{Image Simulation}
\label{subsection: ImageSimulation}

\subsubsection{Calculating Surface Brightness}
\label{subsubsection: CalculatingSurfaceBrightness}

We utilize the lens modeling package \textsc{Lenstronomy}\footnote{\url{https://github.com/lenstronomy/lenstronomy}} \citep{Birrer2018, Birrer2021} to perform the ray-shooting and calculate surface brightness. We will refer to this surface brightness distribution with no PSF convolution or detector effects as the ``synthetic image.'' The calculation is performed on a grid that is 5x5 times the pixel density of the final image. Additionally, the bright regions are further supersampled by a factor of three. We define this region by calculating the image positions and determining their radial distances from the center of the scene. Then we build an annulus with inner radius 0.55 arcsec less than the innermost image position and outer radius 0.55 arcsec greater than the outermost image position. We confirm with visual inspection of a representative sample of strong lens morphologies that this annulus encapsulates bright pixels. Through testing, we observe that this further supersampling improves accuracy for these important pixels where the substructure signal is observed.

\subsubsection{Wide Field Instrument Imaging}
\label{subsubsection: WFI}

\textit{Roman's} Wide Field Instrument (WFI) comprises 18 H4RG-10 (HgCdTe) detectors or Sensor Chip Assemblies (SCAs). Each SCA is 4096x4096 pixels at a pixel scale of 0.11 arcsec pixel$^{-1}$, but due to reference pixels on the outer four rows and columns, the usable detector size is 4088x4088 pixels. Geometric distortion is expected to be less than 2\% across the field of view.\footnote{\href{https://roman-docs.stsci.edu/roman-instruments-home/wfi-imaging-mode-user-guide/wfi-design/description-of-wfi}{Roman User Documentation: Description of WFI}, version dated January 5th, 2024}

\begin{figure*}[htb!]
    \centering
    \includegraphics[width=0.9\textwidth]{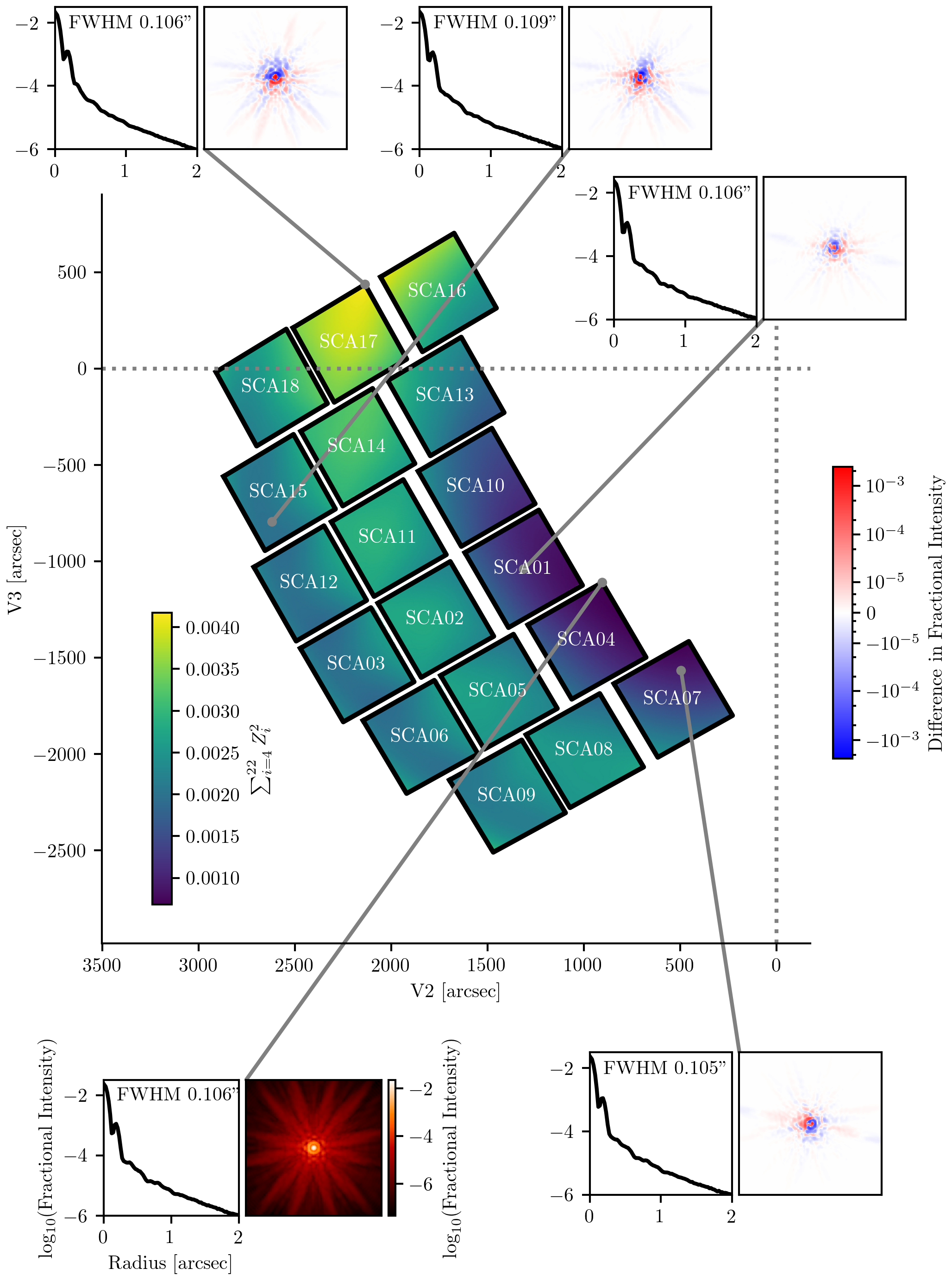}
    \caption{Residuals between PSFs for the F129 filter across the focal plane and PSF radial profiles are plotted in the telescope (V-frame) coordinate system. The intensity of the PSF residual is stretched using an arcsinh scale to show the spatial dependence of these noise sources on the same color scale. The 18 SCAs are colored according to the sum of squared Zernike coefficients $Z_4$ through $Z_{22}$ where smooth spatial variation is achieved by cubic interpolation between the five field points on each SCA provided with the Cycle 9 measurements. The first three Zernike coefficients are left out because they are not relevant for PSFs. The coefficients are plotted for 1.293 microns, the central wavelength of the F129 filter.}
    \label{fig: psfsAcrossFocalPlane}
\end{figure*}

\begin{figure*}[htb!]
    \centering
    \includegraphics[width=\textwidth]{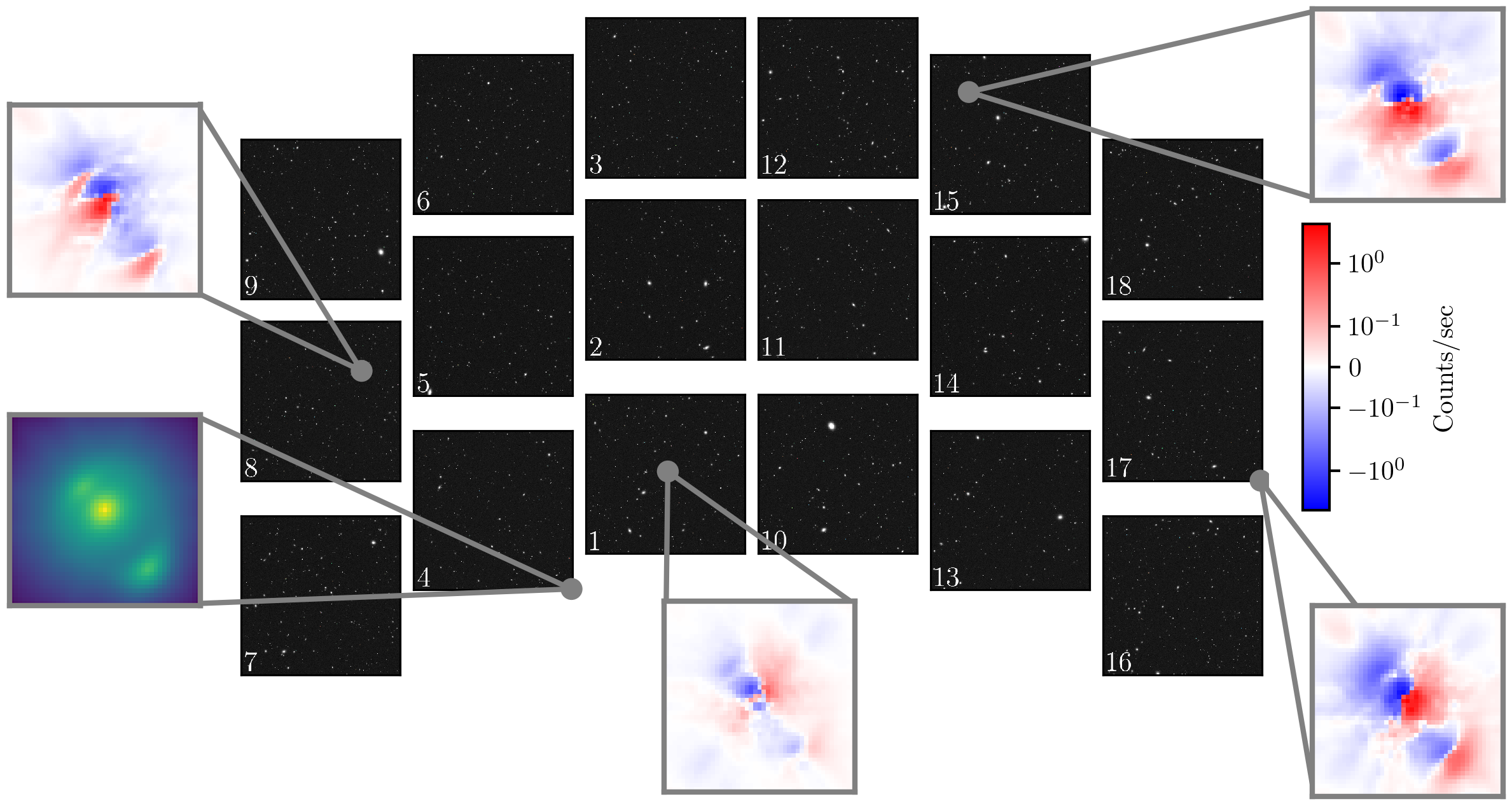}
    \caption{Demonstration of the field-dependence of the PSF across the WFI focal plane. The F184 image of the strong lens with log-scaled intensity is given in the bottom-right corner of SCA04, where wavefront error is minimized (see Figure \ref{fig: psfsAcrossFocalPlane}). We choose the F184 filter for this demonstration because it has the largest FWHM of the anticipated HLWAS filters. Each residual is this image minus an image of the same system that is convolved instead with the PSF at the indicated location on the focal plane. The intensity is arcsinh-stretched to amplify small effects. To isolate the effect of the field-dependent PSF, no detector effects are added. The background is a simulated \textit{Roman} HLWAS image generated using \href{https://github.com/spacetelescope/STScI-STIPS}{\textsc{STIPS}} \citep{Gomez2024}.}
    \label{fig: fieldDependence}
\end{figure*}

The distinctive arrangement of the 18 SCAs minimizes wavefront error across the WFI's wavelength range, given \textit{Roman's} three-mirror anastigmatic optical design  \citep{Pasquale2018}. The latest measurements of optical aberrations are provided with \textit{Roman's} Cycle 9 specification as measurements of Zernike coefficients with Noll ordering for $Z_1$ through $Z_{22}$ at 21 wavelengths from 0.48-2.3 microns and at five positions on each SCA. This field-dependent variation of wavefront error across the WFI focal plane is demonstrated in Figure \ref{fig: psfsAcrossFocalPlane}. Figure \ref{fig: fieldDependence} shows the effect this has on images of strong lenses, which have complex morphologies given the range of possible galaxy shapes and configurations.

\begin{figure*}[htb!]
    \centering
    \plotone{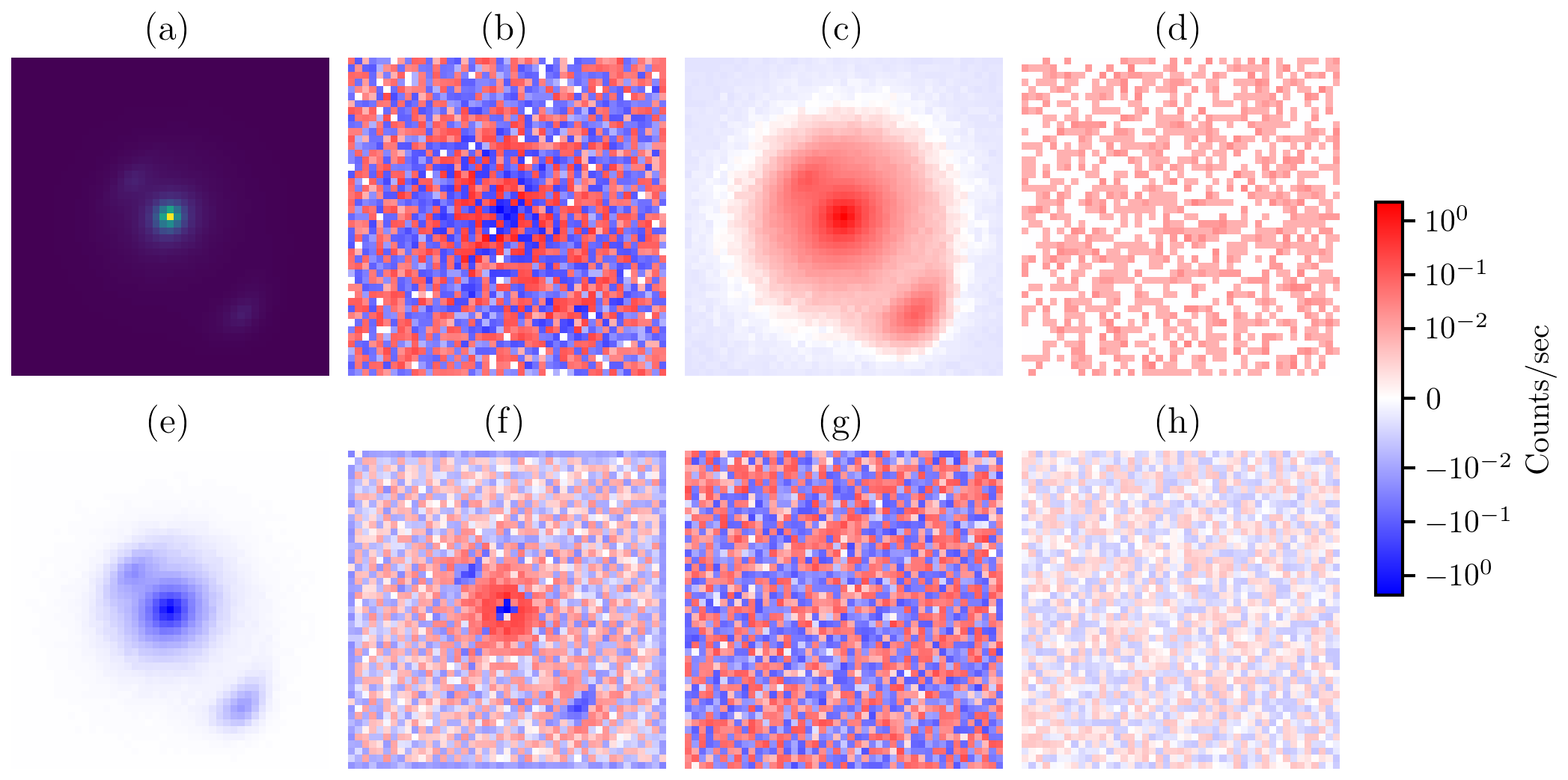}
    \caption{Applying \textit{Roman} WFI detector effects in order to an image of the system in Figure \ref{fig: romanImage}. (a) Synthetic image convolved with PSF and sky background added, (b) Poisson noise, (c) reciprocity failure, (d) dark current, (e) nonlinearity, (f) interpixel capacitance, (g) read noise, and (h) quantization. Intensity is stretched using an arcsinh scale to show the spatial dependence of these effects on the same color scale. Note that the Poisson noise residual has negative values because each pixel value of the synthetic image with sky background is taken to be the expectation value of a Poisson process.\newline\newline}
    \label{fig: noise}
\end{figure*}

We simulate the following WFI detector effects in the order they occur in the actual instrument: Poisson noise, reciprocity failure, dark current, nonlinearity, interpixel capacitance, read noise, and quantization due to an integer number of counts being detected. Figure \ref{fig: noise} shows the scale and morphology of each of these effects by taking the difference between the image before and after applying the effect.

Reciprocity failure is reduced pixel response at low flux levels. The pixel response with reciprocity failure $p_R$ can be estimated in terms of the nominal pixel response $p$ according to the following expression utilized by \textsc{GalSim} \citep{Rowe2015}:
\begin{equation}
    \frac{p_R}{p} = \left(\frac{p}{t}\right)^{\alpha/\log(10)}
\end{equation}
where $\alpha = 6.5 \times 10^{-3}$ and $t$ is the exposure time, $146$ s for the HLWAS. 

Dark current is due to thermal photons, as the detectors are expected to be at an operating temperature of 95 K. Testing of flight candidate detectors suggests that the expected performance is $<0.5 \times 10^{-2}$ electrons pixel$^{-1}$ s$^{-1}$ \citep{Mosby2020}.

Nonlinearity is due to voltage nonlinearity in the charge-to-voltage conversion process, which attenuates brighter pixels. The accumulated charge including the effect of nonlinearity $Q_\textrm{NL}$ in terms of the accumulated charge $Q$ is 
\begin{equation}\label{eq: nonlinearity}
    Q_\textrm{NL} = Q + \beta Q^2
\end{equation}
where the $Q$s are in units of number of electrons and $\beta = -6 \times 10^{-7}$ per electron based on testing of a prototype \textit{Roman} detector \citep{Choi2020}. 

Interpixel capacitance is the effect of charges in neighboring pixels affecting the voltage read at a given pixel. Thus, it can be approximated by convolving the image with a 3x3 kernel representing the signal distribution among neighboring pixels. We use the normalized kernel
\begin{equation}
    \begin{pmatrix}
        0.00200844 & 0.01908014 & 0.00210886 \\
        0.01847761 & 0.91574613 & 0.01827676 \\
        0.00220928 & 0.01988351 & 0.00220928
    \end{pmatrix}
\end{equation}
from measurements of flight candidate detectors in Phase C specifications dated January 8th, 2021.\footnote{\url{https://roman.gsfc.nasa.gov/science/RRI/Roman_WFI_Reference_Information_20210125.pdf}} This convolution introduces a border around the image, which we crop off later.

Finally, read noise is due to errors with reading out pixel values from electrons on the detector. We simulate it using \textsc{GalSim}'s \texttt{roman} module (see Section \ref{subsubsection: CalculatingObservedImage}): Gaussian noise with $\sigma = 8.5$ electrons.

\subsubsection{Calculating Observed Image}
\label{subsubsection: CalculatingObservedImage}

To include the PSF, sky background, and detector effects in our simulated images, we utilize the \texttt{roman} module of the open-source image simulation package \textsc{GalSim}\footnote{\url{https://github.com/GalSim-developers/GalSim}} \citep{Rowe2015, Troxel2021} which can apply realistic telescope effects to astronomical scenes. \textsc{GalSim}'s \texttt{roman} module comprises a suite of methods and attributes based on the design and anticipated performance of \textit{Roman's} instruments. To apply \textit{Roman} detector effects to the synthetic image from \textsc{Lenstronomy}, we first calculate the total flux of the system in counts s$^{-1}$, then import the image as an interpolated image to \textsc{GalSim} with the appropriate flux and pixel scale information. Based on the design reference mission HLWAS, we set the exposure time to 146 s. Then, a detector from the WFI's 18 SCAs and a usable pixel position on that detector are randomly chosen to be the center of the image from among 450 uniformly-spaced positions across the focal plane where PSFs have been pre-generated and cached. Usable pixels are those far enough from the edges such that the entire scene will appear on the detector, accounting for the calibration pixels along the edges. An achromatic PSF defined at the appropriate effective wavelength is then calculated at that particular position and convolved with the image for that band. The PSF and image are supersampled by a factor of five to achieve accurate ray-shooting through substructure, flux calculations, and PSF asymmetry effects while conserving computational resources.

We use \textsc{WebbPSF}\footnote{\url{https://github.com/spacetelescope/webbpsf}} \citep{Perrin2012}, a PSF simulation package for \textit{Roman} from STScI, to generate PSFs. Until empirical PSFs from in-flight observations can be developed, \textsc{WebbPSF} is the optimal choice for accurate PSFs. It is expected that a library of empirical PSFs from across the focal plane will be maintained by the \textit{Roman} Science Operations Center (SOC). These PSFs will be derived from dithered observations of point sources.

The shape and transmission of the PSF varies across \textit{Roman's} wide field of view. By drawing PSFs from across \textit{Roman's} 18 SCAs, we account for these variations. To infer the presence of milli-arcsecond lensing due to dark matter substructure from imaging data, such variations are non-negligible (see Section \ref{subsection: SubstructureSignalAndDetectorEffects}). \textsc{WebbPSF} calculates the optical aberration at a given position by doing cubic interpolation of measurements of Zernike coefficients provided with \textit{Roman's} Cycle 9 specification. \textsc{GalSim}'s \texttt{roman} module has a method \texttt{getPSF} that uses the same aberration data as \textsc{WebbPSF}, but does linear interpolation. The PSF accounts for the expected jitter of 0.012 arcsec per axis and is normalized at the entrance pupil. This normalization accounts for diffractive losses through the instrument, resulting in a field-dependent transmission between 96.9\% and 97.3\% where transmission tends to be greatest at the center of the focal plane array and decreases at the right and left outer edges.

Zodiacal light and thermal background are calculated using count rates per pixel from the latest WFI technical documentation.\footnote{See footnote \ref{footnote: RomanTechnicalDocumentation}.} For zodiacal light, we take 1.5 times the minimum zodiacal light, which is typical at high galactic latitudes. Then, stray light at 10\% zodiacal light is added, though the actual performance is expected to be much better. The total sky background in each band is then the sum of zodiacal light, stray light, and thermal background.

Finally, the \texttt{galsim.roman.allDetectorEffects} method applies all noise and detector effects in the appropriate sequence, as described in the previous section (Section \ref{subsubsection: WFI}). The noise sources included are Poisson noise, dark current, and read noise. The detector effects include reciprocity failure, quantization, nonlinearity, and interpixel capacitance. We do not include the impact of persistence because we are only considering individual exposures and are not simulating a series of previous exposures. The outer three rows/columns of pixels are cropped off to eliminate edge effects due to the interpixel capacitance convolution step. The final image is then converted to units of counts s$^{-1}$. As such, it does not fit perfectly into the data processing levels (Level 0 through 5) of \textit{Roman} data products, but is roughly a Level 2 image prior to the various calibration steps of the exposure processing pipeline since it is a single exposure in units of counts s$^{-1}$. We assume a CCD gain of unity, so this is equivalent to Digital Number (DN) s$^{-1}$.

%-----------------------------------------------------------------------
\section{Results}
\label{section: Results}

\subsection{Survey Simulation Results}
\label{subsection: SurveySimulationResults}

\begin{figure*}[htb!]
    \centering
    \plotone{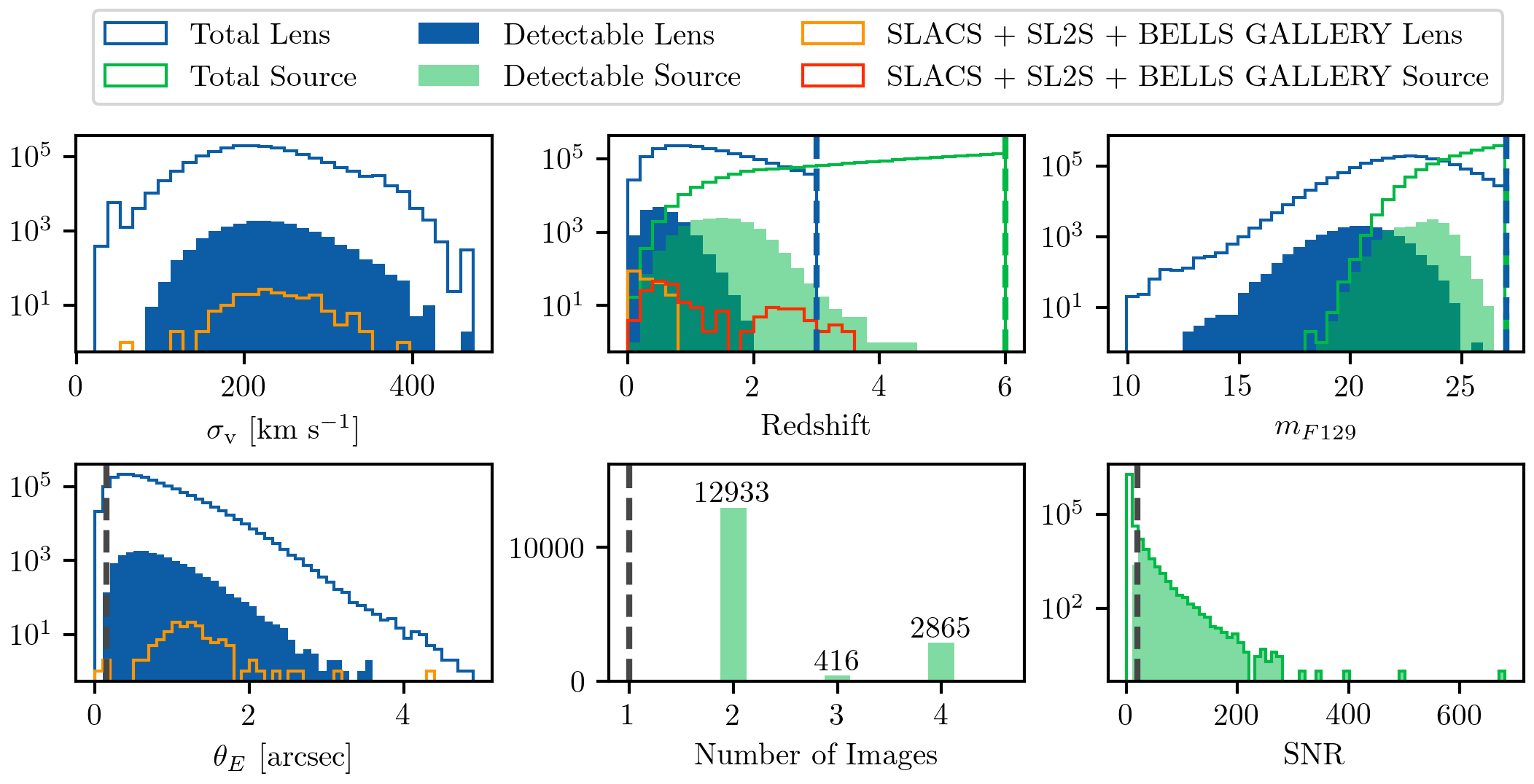}
    \caption{Parameters of the detectable strong lenses from the survey simulation, which is about 16\% of the $\sim$1700 deg$^2$ HLWAS. Top row, from left to right: velocity dispersion of lensing galaxies, redshifts of source and lensing galaxies, magnitudes of lensing and (unlensed) source galaxies. Bottom row, from left to right: Einstein radii, number of images formed, SNRs (as defined in Section \ref{subsubsection: DetectabilityCriteria}). We overplot parameters of strong lenses with \textit{HST} observations from the Sloan Lens ACS Survey \citep[SLACS;][]{Bolton2008}, Strong Lensing in the Legacy Survey (SL2S) project \citep{Sonnenfeld2013a, Sonnenfeld2013b}, and BELLS GALLERY Survey \citep{Shu2016a, Shu2016b}. Note that of the 131 SLACS systems, Einstein radii are available for the 63 grade A SLACS lenses amenable to modeling \citep[Table 5]{Bolton2008}. Colored dotted lines indicate the cutoffs listed in Table \ref{table: surveyParams} applied to the galaxies before determining which ones will form strong lenses. The gray dotted lines indicate cutoffs that are part of the detectability criteria (Section \ref{subsubsection: DetectabilityCriteria}).}
    \label{fig: skypySummary}
\end{figure*}

Assuming a fiducial case of single 146 s exposures, per square degree of the simulated survey area, we find $\sim$95 strong lenses that meet the criteria for detectability. This corresponds to roughly 27 detectable strong lenses per 0.281 deg$^2$ HLWAS exposure and 160,000 in the full 1700 deg$^2$ survey. The parameters of these strong lenses are summarized in Figure \ref{fig: skypySummary}. Of these, 0.28 systems per deg$^2$ or around 500 in the HLWAS have SNR $>$ 200 and are thus amenable to substructure characterization. Assuming coaddition of four 146 s exposures resulting in an effective exposure time of 584 s, we predict upper limits of 510,000 (300 per deg$^2$) detectable and 5600 (3.3 per deg$^2$) characterizable strong lenses in the HLWAS. These estimates based on effective exposure times from image coaddition are upper limits because effective exposure time does not scale linearly with number of exposures. We take SNR $>200$ to be a condition for substructure characterizability based on comparison with \citet{Despali2022}, which finds that substructure detectability depends primarily on SNR and takes systems with median SNR per pixel of 3.5 to 4.5 as examples of systems amenable to substructure characterization. A cutoff at overall SNR $=200$ selects systems with a median SNR per pixel of $3.7 \pm 0.5$.

Compared to \textit{HST's} population of strong lenses represented by the SLACS, SL2S, and BELLS GALLERY samples, the predicted \textit{Roman} population has more systems with smaller Einstein radii and higher redshifts. The smaller Einstein radii are likely due to selection effects. For example, the SLACS systems were identified by SDSS spectroscopic measurements of galaxies where emission lines from two distinct redshifts were identified \citep{Bolton2006}. Owing to selection criteria to maximize the number of confirmed systems, candidates with larger velocity dispersions and therefore larger Einstein radii were selected for \textit{HST} follow-up. \textit{Roman's} population will extend to much higher redshifts due to \textit{Roman's} deep limiting magnitudes and NIR coverage. In contrast, strong lenses identified by ground-based surveys tend to be relatively bright and nearby.

Our results broadly agree with predictions from previous work for strong lens detections in \textit{Roman's} HLWAS. \citet{Weiner2020} uses the strong lens population simulation code \texttt{LensPop}\footnote{\url{https://github.com/tcollett/LensPop}} \citep{Collett2015} to predict 16,778 detectable systems in the HLWAS, which they take to be 2000 deg$^2$, coming out to roughly eight detectable strong lenses per deg$^2$. We improve upon this work by using a more detailed galaxy simulation from \textsc{SLSim} and \textsc{SkyPy} and updated \textit{Roman} HLWAS and detector parameters. In particular, as noted in \citet{Holloway2023}, they use a zero-point magnitude of 23.9 for F129, which corresponds to a less sensitive detector than current measurements indicate (26.31 for SCA13 to 26.47 for SCA18). While the parameters of our lensing galaxies are in good agreement, as they find lens redshifts with mean $0.6 \pm 0.3$, velocity dispersions with mean $219 \pm 50$ km s$^{-1}$ and we find $0.57 \pm 0.27$ and $229 \pm 51$ km s$^{-1}$, respectively, they predict a dimmer and more distant population of source galaxies (source magnitudes $24.8 \pm 1.6$ and source redshifts $1.9 \pm 0.9$ compared to $23.1 \pm 1.1$ and $1.5 \pm 0.5$, respectively).

\citet{Holloway2023} predicts $\sim$52 strong lenses per deg$^2$ in the HLWAS. They note that their pipeline is optimized for surveys that are smaller and deeper than \textit{Roman's} HLWAS due to their choice of galaxy catalog, and they expect that their extrapolation will underestimate the number of detectable systems due to a dearth of higher-mass ($>10^{11.5}$ M$_\odot$) lensing galaxies. Our yield is roughly double, likely due to our inclusion of massive early-type elliptical galaxies which comprise a substantial fraction of the lensing galaxies of detectable systems. \citet{Ferrami2024} uses an analytical model but similar detectability criteria and survey and instrument parameters to \citet{Holloway2023} and predicts 4.7-10.0 systems per deg$^2$ depending on the choice of velocity dispersion function, where they note that their model is sensitive to this poorly-constrained function. We use their \textsc{GaLeSS}\footnote{\url{https://github.com/Ferr013/GALESS}} code \citep{Ferrami2024} to produce parameter distributions for detectable systems in the HLWAS, and find that our results for velocity dispersions, redshifts, and Einstein radii are in good agreement.

\citet{Daylan2023} focuses on the potential for \textit{Roman} to constrain substructure down to $10^7$ M$_\odot$. A survey strategy better optimized for this science case would utilize shorter wavelength filters and a longer exposure time than the anticipated HLWAS to achieve higher SNR. For strong lenses amenable to substructure characterization, they require a separation of 0.8 arcsec between at least two images and magnified source magnitude $m_\textrm{F087} < 21$. In a simulated survey using the F087 and F106 bands with 3600 s exposures, they find approximately 500 strong lenses that meet these criteria in 1700 deg$^2$, or 0.29 per deg$^2$. In this work, we use SNR $>$ 200 as a threshold for characterizability and find 0.28 systems per deg$^2$ that meet this criterion with 146 s exposures.

\subsection{Image Simulation Results}
\label{subsection: ImageSimulationResults}

\begin{figure*}[htb!]
    \centering
    \plotone{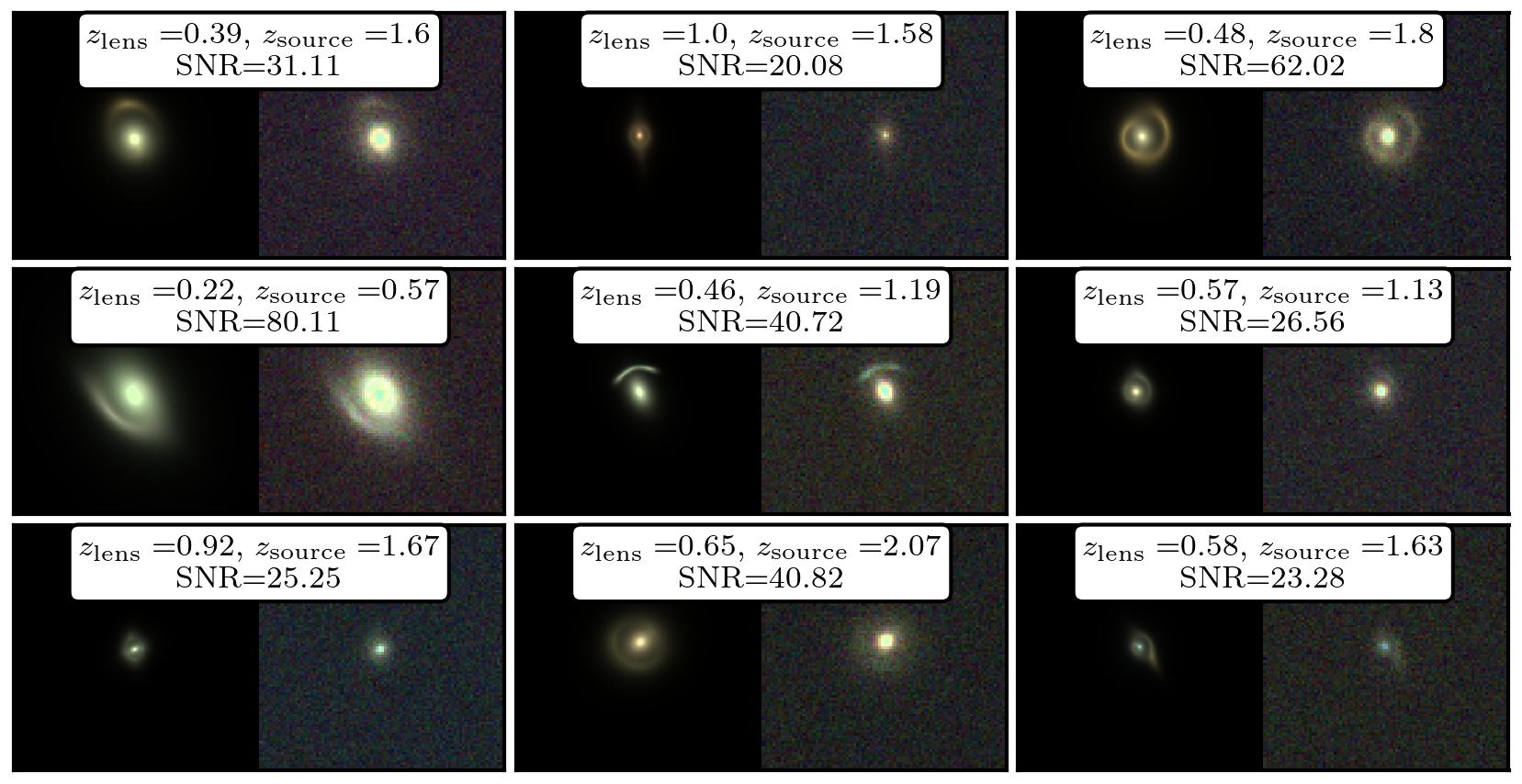}
    \caption{A subset of the simulated images. The synthetic images are on the left, while the corresponding images with sky background and WFI detector effects are on the right. The cutouts are 10.01 arcsec on a side at the native pixel scale of 0.11 arcsec pixel$^{-1}$.}
    \label{fig: mosaic}
\end{figure*}

Our dataset comprises 16,214 images and is available on Zenodo \citep{Wedig2024}. Each image is 10.01 arcsec or 91 pixels on a side. Figure \ref{fig: mosaic} shows a random subset.

The survey simulation takes 28 minutes per deg$^2$ of the simulated survey area. The image simulation takes 4.5 seconds per image, where the primary contributor (95\%) is performing the ray-shooting through the subhalos to generate a supersampled image. The generation of the entire dataset took 99 hours and was performed on a machine with 32 AMD Opteron 6328 3.20 GHz CPUs and 128 GB of RAM. The pipeline uses Python's \texttt{multiprocessing} module to execute many processes simultaneously in parallel on high core-count processors, reducing total execution time roughly linearly by the number of available logical processors.

%-----------------------------------------------------------------------
\section{Discussion}
\label{section: Discussion}

\subsection{Comparing Lower-mass Substructure Signal with WFI Systematics}
\label{subsection: SubstructureSignalAndDetectorEffects}

We quantify the effects of the variation of the PSF and zero-point magnitudes across \textit{Roman's} WFI on the signature of dark matter substructure. In other words, how do \textit{Roman} systematics affect the substructure signal? The PSF can be measured empirically and therefore accounted for when inferring substructure in strong lenses (e.g., by including it in a forward model). Poisson noise and read noise are pixel-scale fluctuations in flux degenerate with the substructure signal, but are random.

The effects of \textit{Roman} systematics on weak lensing measurements have been investigated in depth in the literature \citep[e.g.,][]{Troxel2021, Yamamoto2024, Laliotis2024}, but we believe we are the first to consider them in the context of studying dark matter substructure with strong lenses. Here, we compare the variation of subhalo populations with the variation of the PSF and zero-point, i.e., the position on a particular detector where a system is observed. 

We first select high-SNR systems from the HLWAS simulation, requiring SNR $>200$. We generate unique subhalo populations for these systems, varying the subhalo mass range. The default subhalo mass range for the dataset is $10^6-10^{10}$ M$_\odot$, but we also consider the ranges $10^{7}-10^{10}$, $10^{8}-10^{10}$, and $10^9-10^{10}$ M$_\odot$ to study the effect of lower-mass subhalos. For each system, the higher-mass subhalos stay in the same positions across the three subhalo populations, and differences in mass are accounted for with negative mass sheets. Since we are interested in quantifying the milli-lensing signature of lower-mass subhalos, we utilize the F106 filter which has the smallest PSF FWHM of the HLWAS filters. We assume an effective exposure time of 438 s based on the three dither positions for the F106 filter in the design reference mission HLWAS. However, the results do not change qualitatively with longer exposure times. To isolate the substructure signal and allow for a direct comparison with the effect of PSF variation, no sky background or detector effects are added and the light of the lensing galaxy is removed.

Free-streaming effects of WDM are expected to suppress the formation of subhalos below a certain mass scale \citep{Schneider2013}, often given by the half-mode mass $m_\textrm{hm}$ defined by where the square root of the ratio of the power spectra of WDM to CDM is one-half. Assuming all WDM are thermal relics, the half-mode mass can be converted to the thermal relic mass $m_\textrm{DM}$ \citep{Schneider2012, Gilman2020}:
\begin{equation}
    m_\textrm{hm} = 3\times10^8\left(\frac{m_\textrm{DM}}{3.3\,\textrm{keV}}\right)^{-3.33}\, \textrm{M}_\odot.
\end{equation}
Using \textit{HST} imaging of strong lenses, \citet{Birrer2017} find $m_\textrm{DM} > 2$ keV ($m_\textrm{hm} > 10^{9.2}$ M$_\odot$) at $2\sigma$ and \citet{Gilman2020} find $m_\textrm{DM} > 5.2$ keV ($m_\textrm{hm} > 10^{7.8}$ M$_\odot$) at $2\sigma$. \citet{Keeley2024} use \textit{JWST} MIRI imaging of 31 strongly-lensed quasars to achieve a constraint of $m_\textrm{DM} > 6.1$ keV ($m_\textrm{hm} > 10^{7.6}$ M$_\odot$) at posterior odds 10:1. Other probes constraining the WDM thermal relic mass include the Lyman-$\alpha$ forest flux power spectra where \citet{Irsic2024} find $m_\textrm{DM} > 5.7$ keV ($m_\textrm{hm} > 10^{7.7}$ M$_\odot$) at $2\sigma$, and Milky Way satellite counts where \citet{Newton2021} find $m_\textrm{DM} \geq 3.99$ keV ($m_\textrm{hm} > 10^{8.2}$ M$_\odot$) at $2\sigma$ and \citet{Nadler2021} find $m_\textrm{DM} > 6.5$ keV ($m_\textrm{hm} > 10^{7.5}$ M$_\odot$) at $2\sigma$. Hence, we consider truncations of lower-mass subhalos at $10^7$, $10^8$, and $10^9$ M$_\odot$ to represent signals that are beyond, at, and within current detection capabilities. 

\begin{figure*}[htb!]
    \centering
    \includegraphics[width=0.75\textwidth]{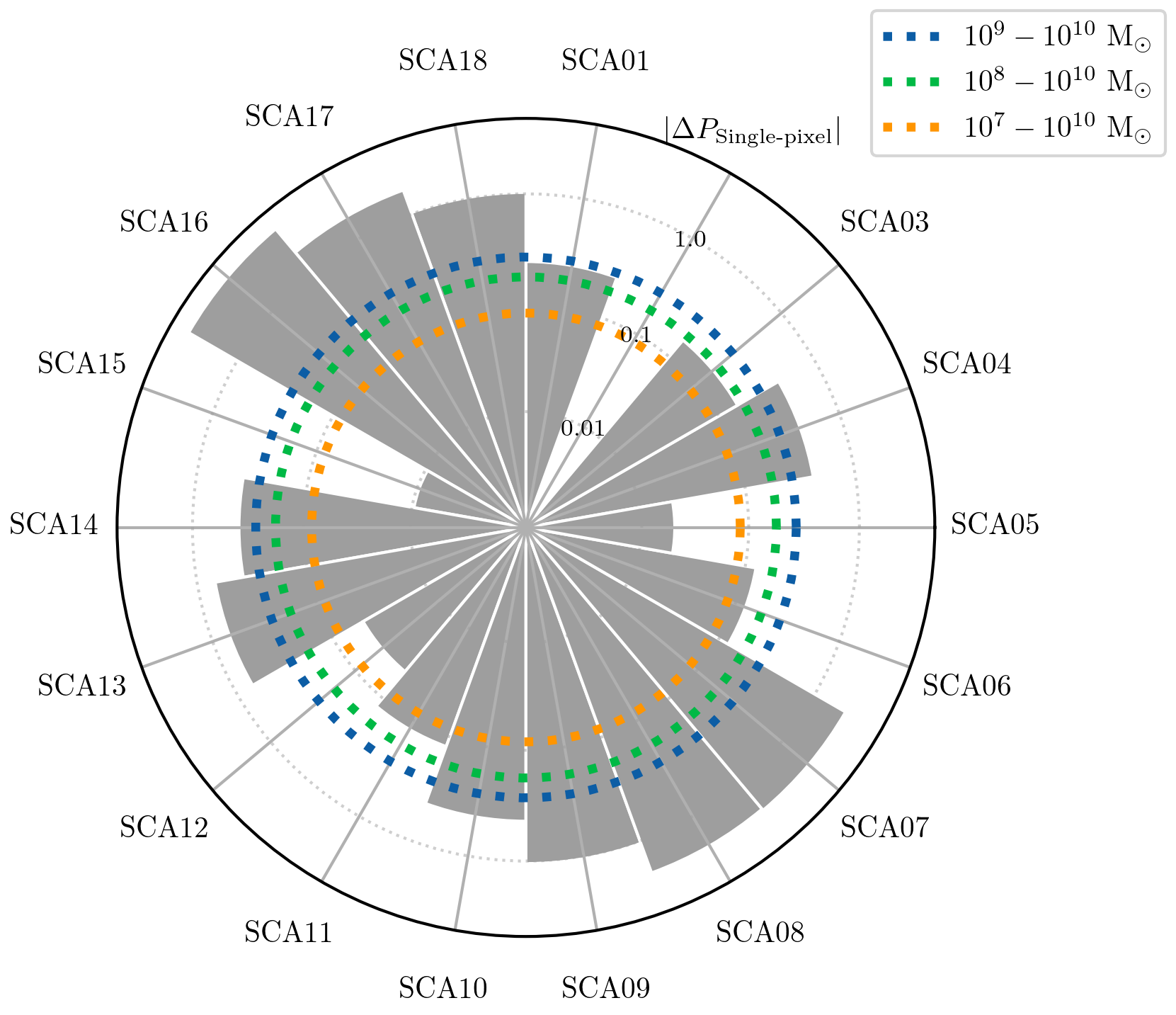}
    \caption{Comparison of the radially-averaged power spectra at the single-pixel scale when varying the PSF and the lower-mass limit of the subhalo population. The radial bars and dotted circles represent the absolute value of the deviation at the single-pixel scale of the power spectrum of an image from the control case: CDM subhalos drawn from the mass range $10^6-10^{10}$ M$_\odot$ and convolved with a PSF near the center of the focal plane (center of SCA02). The radial direction is on a logarithmic scale. For most SCAs, the systematic PSF variation exceeds, and therefore masks the effect of lower-mass subhalos. For massive lensing galaxies ($>10^{13}$ M$_\odot$) which are expected to have more subhalos, the single-pixel power due to subhalos roughly doubles across all subhalo mass ranges.}
    \label{fig: powerSpectraPolar}
\end{figure*}

The lensing signal due to subhalos is at or below the milli-arcsecond scale, so as a simple proxy of this signal, we consider the power spectra of images at small angular scales. Images of strong lenses with more substructure should have more power in their radially-averaged power spectra at the single-pixel scale due to milli-lensing. We use the \texttt{lenstronomy.Util.correlation.power\_spectrum\_1d} method to calculate the radially-averaged power spectra of the images. 

Figure \ref{fig: powerSpectraPolar} compares the effects of varying subhalo populations and varying focal plane position at the single-pixel scale. We find that across most of the WFI focal plane, focal plane position skews single-pixel power more than the effect of lower-mass subhalos. In other words, the effect of subhalos that may be less abundant in the WDM case is approximately at or below the scale of the effect of the variation of the PSF across most of the focal plane. This effect is most pronounced at the edges of the focal plane (SCAs 7, 8, 9, 16, 17, and 18) as we are considering variations with respect to the PSF at the center of the centrally-located SCA02. Therefore, an analysis pipeline doing substructure inference on \textit{Roman} strong lenses, particularly for systems that happen to be found serendipitously in wide surveys, should take into account the PSF at the position where each system happens to be detected. The shape of the PSF at any given position can be modeled by, e.g., \textsc{WebbPSF}, and will be measured once the instrument is in flight.

\subsection{Single-subhalo Detection Across the WFI}
\label{subsection: LowestDetectableSubhaloMass}

In this section, we consider the detection of individual subhalos. Single-subhalo detections have been used to place constraints on WDM \citep[e.g.,][]{Vegetti2018, Ritondale2019}, the CDM substructure mass fraction \citep[e.g.,][]{Vegetti2010, Vegetti2012, Vegetti2014}, and subhalo density profiles \citep[e.g.,][]{Despali2024}. WFI systematics across the focal plane will also have non-negligible effects on the detection threshold for a single massive subhalo. 

However, the HLWAS will not have a sufficient exposure time to permit the detection of single subhalos with masses below around $10^{10}$ M$_\odot$ \citep{Daylan2023}. Therefore, in this section we consider a follow-up survey of high-SNR strong lenses with sufficient depth to detect lower-mass single subhalos. \textit{Roman's} planned High Latitude Time Domain Survey (HLTDS) will provide a sample of such systems. Its primary objective is to observe type Ia supernovae to study dark energy. While the details of this survey strategy have not yet been finalized, it is expected to comprise wide (19.04 deg$^2$) and deep (4.20 deg$^2$) imaging survey components with the F062/F087/F106/F129 and F106/F129/F158/F184 filters, respectively \citep{Rose2021b}. To estimate the strong lens yields in these components, we use our survey simulation code described in Section \ref{subsection: SurveySimulation} to simulate small patches ($\sim$ 1 deg$^2$), then extrapolate and relax the magnitude cuts to allow for dimmer systems. We predict upper limits of roughly 44,000 (2300 per deg$^2$) detectable systems in the wide component using an effective exposure time of 12,500 s and 25,000 (6000 per deg$^2$) in the deep component (37,500 s). Of these, we predict 2100 and 1200 systems, respectively, will be amenable to substructure characterization due to their high SNR.

For the deep observations we consider in this section, we utilize the F087 filter, which has a smaller PSF FWHM than those of the HLWAS filters. The F062 filter has the smallest PSF FWHM of the eight WFI filters, but at the cost of lower flux due to the spectral energy distributions of the galaxies comprising detectable systems. We use an exposure time of 12,500 s. We select the highest-SNR systems from the HLWAS survey simulation:  bright and well-aligned systems that are ideal for substructure characterization. Each system is imaged at 16 uniformly spaced positions on each of the 18 detectors. At each position, we simulate an exposure of the system as-is, then simulate an exposure with a single subhalo placed at an image position. Subhalo detections are the most likely for subhalos near image positions where SNR is highest \citep{Despali2022}. The subhalo is parameterized with an NFW density profile where we hold the concentration constant at 10 instead of sampling from a mass-concentration model so as to isolate the effect of PSF variation. Detector effects are computed for the system without the additional subhalo and re-used across all copies of that system. We do not apply the effect of nonlinearity because the way that it is modeled is inaccurate for the long exposure times considered here. For large accumulated charge $Q$ the quadratic term dominates and the nonlinearity becomes negative for $Q>\frac53\times10^6$ (see Equation \ref{eq: nonlinearity}). 

\begin{figure}[htb!]
    \centering
    \plotone{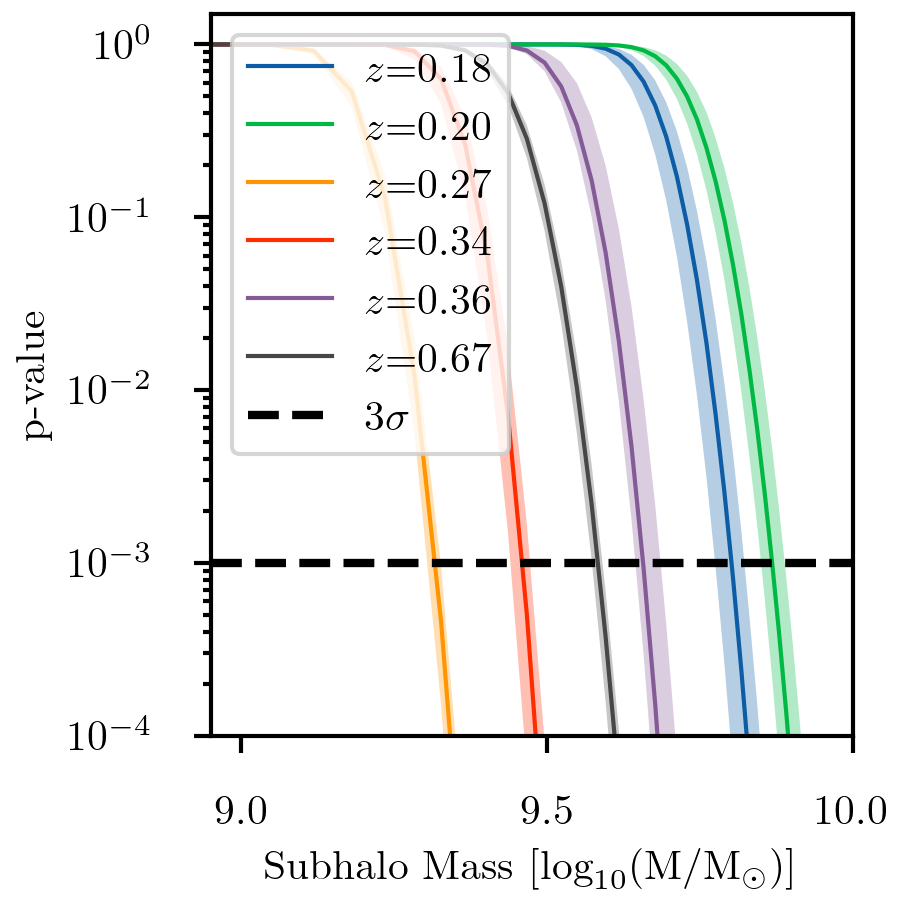}
    \caption{Demonstration of the increase in detection significance of single subhalos as their masses increase. Each line represents a random instance of a high-SNR system with a single subhalo of varying mass located at an image position. The shaded region represents variation in the PSF at 288 uniformly spaced positions across the focal plane, and the solid line is the median. This distribution of p-values at which the subhalo becomes detectable at $3\sigma$ confidence is represented in Figure \ref{fig: lowestDetectableSubhaloMass} by showing the mass at which the subhalo becomes detectable as a function of the position on the focal plane where it was imaged. We label these samples by the redshift of the subhalo, which is also the redshift of the lensing galaxy. We note that subhalo detectability is a complicated function of SNR and observational parameters and does not necessarily correlate with redshift.}
    \label{fig: pvals}
\end{figure}

To characterize the effect of field-dependent \textit{Roman} systematics on single-subhalo detection, we consider the least massive subhalo that in principle could be detected across the WFI focal plane. To determine whether a subhalo is detectable, we use a chi-squared test statistic. We compute 
\begin{equation}\label{eq: chi2}
    \chi^2 = \sum_{i} \frac{\left(A_{i} - B_{i}\right)^2}{B_{i}}
\end{equation}
between each image with a subhalo ($A$) and without a subhalo ($B$), where the summation is over pixels where the SNR per pixel $>1$ in the HLWAS, i.e., the region where the lensed emission has the highest SNR (see Section \ref{subsubsection: DetectabilityCriteria}). We consider subhalos in the mass range $10^8$ to $10^{10}$ M$_\odot$. Subhalos less massive than $10^8$ M$_\odot$, even with high concentrations and excellent alignment with bright images, are unlikely to be detected with sufficient significance as angular resolution sets a lower limit. Subhalos more massive than $10^{10}$ M$_\odot$ are likely to host luminous dwarf galaxies and therefore be directly detectable in the imaging data \citep{Nierenberg2013a}. We add a subhalo, calculate the exposures with and without it, and evaluate Equation \ref{eq: chi2} to quantify the perturbation due to the subhalo. At a certain mass, the subhalo causes enough deflection for $\chi^2$ to exceed a statistical significance of $3\sigma$ by an ideal detection algorithm. Such a detection algorithm is ideal because it attributes every difference between the images of the systems with and without the subhalo to the presence of that subhalo. Then, the mass is incrementally increased and this process is repeated, and once the upper mass limit is reached, the focal plane position is shifted and the entire process repeats. Figure \ref{fig: pvals} illustrates that as the mass of the added subhalo increases and its lensing effect becomes more significant, the $\chi^2$ between the images with and without the subhalo increases and therefore the detection significance increases.

\begin{figure*}[htb!]
    \centering
    \includegraphics[width=\textwidth]{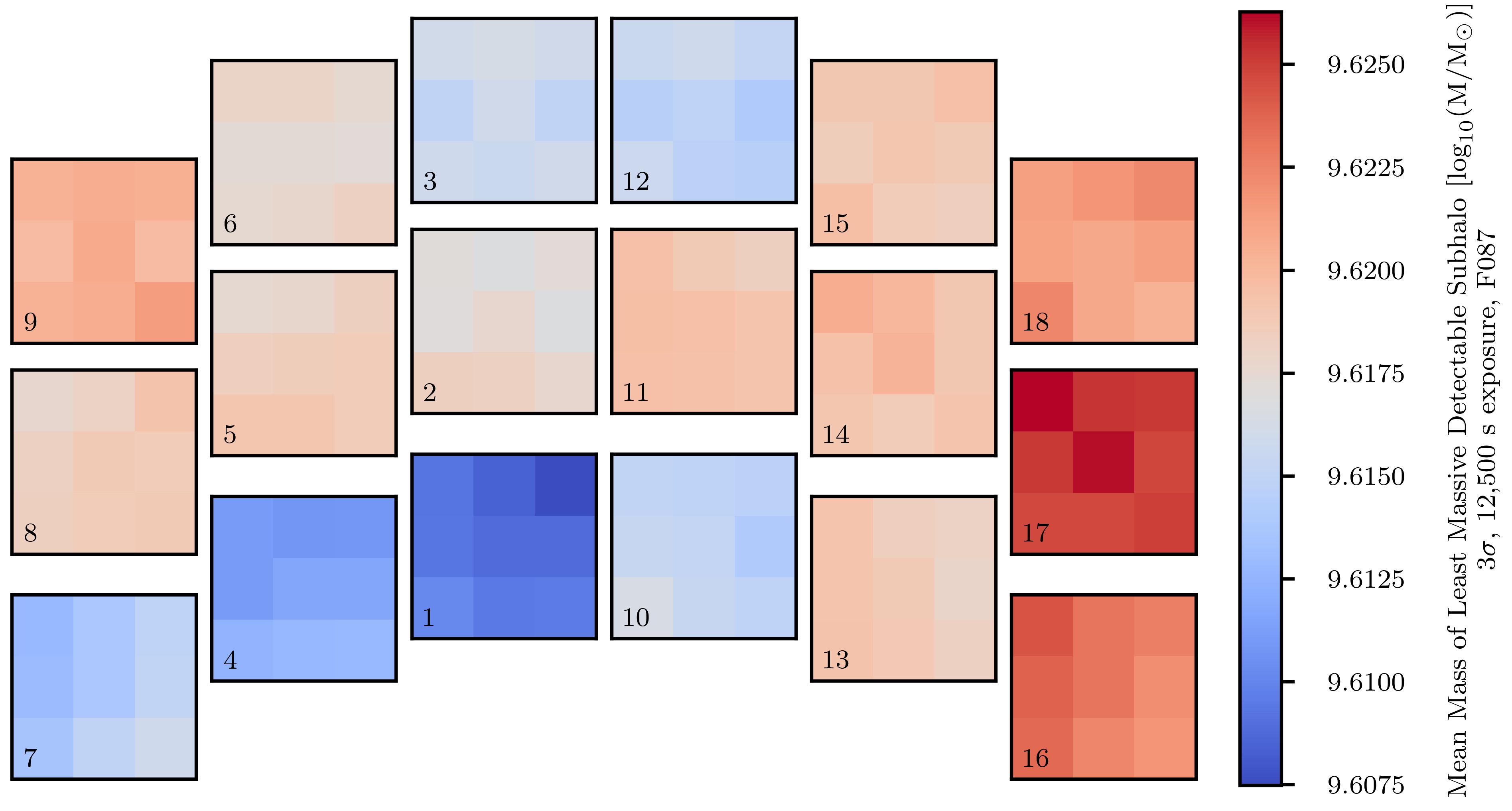}
    \caption{Mean mass of least massive detectable subhalo in high-SNR strong lenses across the WFI focal plane at $3\sigma$ confidence by an ideal detection algorithm with deep \textit{Roman} observations. Each subhalo is placed at an image position, representing the best-case scenario position for being detectable. The PSF is taken at the corresponding SCA position of the center of each subdivision. While the variation across the focal plane of the mass of the least massive detectable subhalo is more strongly affected by the variation in sensitivity across SCAs, when holding zero-point constant as is shown here, it is primarily a function of wavefront error.}
    \label{fig: lowestDetectableSubhaloMass}
\end{figure*}

Figure \ref{fig: lowestDetectableSubhaloMass} shows how focal plane position affects the mass of the least massive detectable subhalo. At each position, we take the mean across the high-SNR systems to average over morphology, brightness, and mass distribution.

We find that the mean mass of the least massive detectable subhalo tracks with wavefront error (see Figure \ref{fig: psfsAcrossFocalPlane}). The average spread across the focal plane is about 5\%, but is as much as 25\% for systems that are particularly sensitive to wavefront error. While this effect is small compared to uncertainties in inferred subhalo mass due to lens modeling \citep{He2023}, it is a well-characterized systematic that can thus be accounted for.

This result is not an estimation of the single-subhalo sensitivity of \textit{Roman} across the WFI, which we plan to address in future work. Here, we isolate the effect of wavefront error on single-subhalo detection by considering only massive, dark subhalos ($10^8$ to $10^{10}$ M$_\odot$) with fixed concentration. We also do not include the variation of zero-point magnitude across SCAs. When including this variation, we observe that mean mass of the least massive detectable subhalo tracks with it instead as the boost in SNR due to a more sensitive SCA has a greater effect on subhalo detectability than wavefront error. However, individual detector sensitivity can be accounted for in targeted observations by choosing a particular SCA to observe with and an exposure time to achieve a particular SNR. We also note that coaddition of exposures will increase single-subhalo sensitivity, but we do not simulate this here.

\subsection{Comparing \textit{Roman} Single-subhalo Detection with \textit{HST}}
\label{subsection: subhaloDetectionCompare}

To date, only a handful of single subhalos have been detected in strong lenses, e.g., one in SDSSJ0946+1006 \citep{Vegetti2010}, one in JVASB1938+66 \citep{Vegetti2012}, one in SDP.81 \citep{Hezaveh2016}, and five candidates\footnote{One of these candidates, SDSSJ0946+1006, was first reported in \citet{Vegetti2010}.} reported in \citet{Nightingale2024}. The first two systems, in particular, have been revisited in a number of papers \citep[e.g.,][]{Despali2018, Minor2021, Sengul2022, Ballard2024, Despali2024, Minor2024}. It has been shown that they are more likely to be line-of-sight halos \citep{Despali2018, Sengul2022}. Five of these systems were discovered solely with optical space telescope photometry (\textit{HST}) in the SLACS and BELLS GALLERY surveys. Subhalos with sufficient mass to be detectable with current instruments, which are limited primarily by angular resolution, are expected to be uncommon: \citet{ORiordan2023} estimate single subhalo detections in $\sim$1\% of \textit{Euclid's} strong lenses.

We consider the single-subhalo detections that might be possible with \textit{Roman} with various survey strategies. We conduct a similar simulation to that in the previous section, using effective exposure times for the HLWAS and HLTDS Wide and Deep components. We fix the PSF to the center of SCA02 and vary the concentration of subhalos by sampling from the mass-concentration model described in Section \ref{subsubsection: CalculatingSurfaceBrightness}. We sample from this realistic mass and redshift-dependent concentration model because not accounting for this spread in concentration can skew subhalo detectability by up to an order of magnitude \citep{Amorisco2022}. To draw a representative number of samples from this model, we generate 1000 subhalos for each mass bin. While the most massive subhalos are likely to have luminous counterparts, we extend the upper mass range to $10^{12}$ M$_\odot$ to compare with \textit{HST} detection candidates. 

\begin{figure}[htb!]
    \centering
    \includegraphics[width=0.45\textwidth]{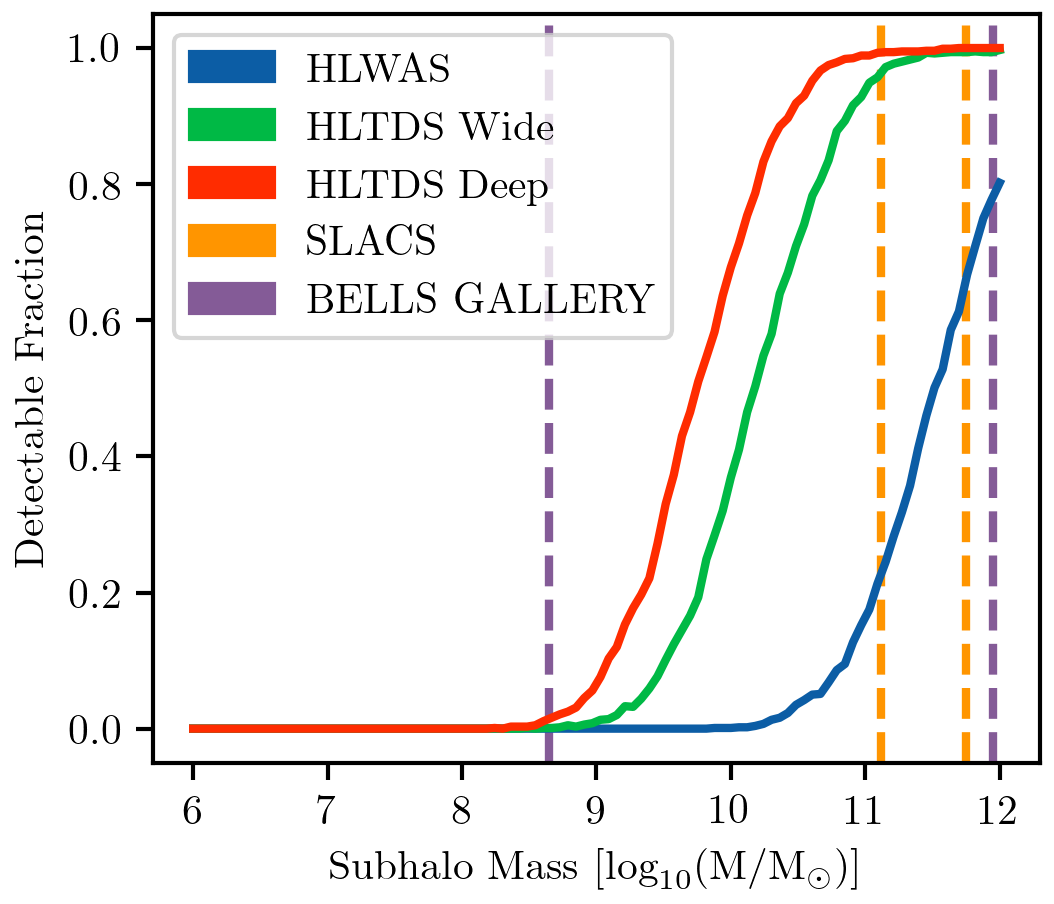}
    \caption{Comparing the distribution of \textit{HST} single subhalo detection candidates reported in \citet{Nightingale2024} from the SLACS and BELLS GALLERY samples with the sensitivity of \textit{Roman} to single subhalos, including the effect of concentration scatter with mass and redshift. While the short exposure times of the HLWAS will not achieve sufficient SNR to probe subhalos below $10^{10}$ M$_\odot$, deep observations are expected to match \textit{HST's} sensitivity and will improve with coaddition. With \textit{Roman's} wide field of view, subhalo searches can be conducted on orders of magnitude more systems.}
    \label{fig: subhaloDetectionCompare}
\end{figure}

In Figure \ref{fig: subhaloDetectionCompare}, we compare \textit{HST} subhalo detection candidates from \citet{Nightingale2024} with those possible with \textit{Roman}. They search for subhalos in a sample of the 54 best systems from the SLACS and BELLS GALLERY samples. Here we choose to plot four of the five candidates, removing the one in SDSSJ0946+1006 due to subsequent investigations that suggest it is more likely to be a line-of-sight halo. We consider the design reference mission HLWAS and HLTDS. We reiterate that even for the highest-SNR systems in the HLWAS, we do not expect to be able to detect single subhalos below $10^{10}$ M$_\odot$ due to the anticipated exposure times not being long enough to achieve sufficient SNR.

\textit{Roman's} angular resolution will be comparable to \textit{HST's}, so we expect its sensitivity to single subhalos to be similar. The least massive subhalo that \textit{HST} is expected to be able to detect is around $10^9$-$10^{9.5}$ M$_\odot$, assuming the pixel scale of around 0.04 arcsec pixel$^{-1}$ of its WFC3 UVIS channel \citep{Amorisco2022, Despali2022}. Based on the image coaddition software \textsc{PyImcom}\footnote{\url{https://github.com/kailicao/pyimcom}} \citep{Cao2024} that is expected to be used, we expect \textit{Roman's} coadd pixel scale to be 0.039 arcsec pixel$^{-1}$ \citep{OpenUniverse2025}.

We reiterate that the collective effect of subhalos below the single-subhalo detectability threshold as described by the normalization of the subhalo mass function, where deviations from the predictions of $\Lambda$CDM are expected, can be probed statistically \citep{CyrRacine2016, WagnerCarena2024, Tsang2024}.

\newpage

\subsection{Complementarity with Euclid}
\label{subsection: Euclid}

\textit{Euclid's} Wide Survey is predicted to observe around 170,000 strong lenses \citep{Collett2015}, and a preliminary search by \citet{AcevedoBarroso2024} suggests that this prediction is still accurate. The Wide Survey is wider than \textit{Roman's} HLWAS, spanning 14,000 deg$^2$ to a depth of AB magnitude 24.5 at visible wavelengths, but the HLWAS is expected to reach a depth of AB magnitude 26.7 in the NIR. The fields are expected to overlap, enabling \textit{Euclid's} coverage to complement \textit{Roman's}. This will enable forced photometry and improved photometric redshifts \citep{Rose2021a}. \citet{Holloway2023} predict that 22\% of the detectable systems in the Wide Survey, primarily those with dusty, lower-redshift source galaxies, will only be detected by \textit{Euclid's} NISP instrument, despite its poorer angular resolution compared to its VIS instrument. \textit{Roman} HLWAS imagery of these systems will achieve higher SNR and angular resolution. For substructure inference, \citet{ORiordan2023} place \textit{Euclid's} lowest detectable subhalo mass at $10^{8.8\pm0.2}$ M$_\odot$ at $3\sigma$. While we leave deriving this limit for \textit{Roman} to future work, assuming that \textit{Roman's} subhalo detection capabilities will be similar to those of \textit{HST}, \textit{Roman's} performance will be similar. Substructure detection and strong lens modeling in general benefit from multi-band imaging as it enables better vetting and characterization of the multiple images.

\subsection{Future Work}
\label{subsection: FutureWork}

The current version of our simulation pipeline has several limitations that we plan to address in future versions. First, we will implement more realistic source and lensing galaxies, using real galaxies from the \textit{HST} COSMOS Survey \citep{Koekemoer2007} instead of S\'ersic profiles. In particular, this will account for irregular galaxy morphologies that are expected at higher redshifts. This will likely boost \textit{Roman's} predicted substructure yield, as sources with concentrated bright features, such as face-on spiral galaxies, are more sensitive to substructure \citep{Ritondale2019, Hughes2024}. We plan to quantify the effect of galaxy morphology on substructure inference in a forthcoming paper (Karthik et al., in prep.).

Additionally, we will model line-of-sight halos. Line-of-sight halos are especially important for \textit{Roman's} population of strong lenses, which extend to much higher redshifts than the well-studied populations observed by \textit{HST} and, therefore, the light from the source traverses a larger volume of line-of-sight halos. Currently, we approximate the effect of massive, distant line-of-sight halos with shear and external convergence terms in the main mass model. \textsc{pyHalo} supports the generation of realistic populations of line-of-sight halos, so we expect the implementation to be seamless. However, assessing their impact on milli-arcsecond lensing in \textit{Roman} images is beyond the scope of this paper.

This work has considered only single exposures, and we leave coaddition to future work. \textit{Roman} is undersampled blueward of about 1.2 microns, so coaddition will enable the recovery of well-sampled images. For the purposes of studying strong lenses, coaddition will improve the angular resolution and SNR, thereby increasing the detectability of strong lenses and their subhalos, but will introduce correlated noise and some deviation from the target PSF. Evaluating these effects on the search for substructure with \textit{Roman's} strong lenses are beyond the scope of this work and will be explored in future work.

While we account for the variation of the PSF across the focal plane array, we do not account for the variation of the PSF across a given 10x10 arcsec scene. We generate the PSF at the center of the scene and convolve the entire image with it. The variation of the PSF at this scale is safe to ignore for this work with galaxy-galaxy strong lenses since the Einstein radii of these systems are $<2.5$ arcsec. However, for strong lensing by galaxy clusters where Einstein radii can be at the arcminute scale, we expect the variation of the PSF across the scene to be a non-negligible effect. 

In future versions of \textsc{mejiro}, we plan to utilize the \textit{Roman} WFI image simulation package \textsc{romanisim}\footnote{\url{https://github.com/spacetelescope/romanisim}} maintained by STScI that is currently under active development and not yet formally validated. \textsc{romanisim} is a \textsc{GalSim}-based package that will be the state-of-the-art for WFI image simulation. Like \textsc{romanisim}, \textsc{mejiro}'s image simulation module is based on \textsc{GalSim} and uses PSFs generated with \textsc{WebbPSF}. \textsc{romanisim}'s improvements include simulating individual reads and resultants according to multi-accumulation tables (``L0'' and ``L1'' data products according to \textit{Roman's} WFI data processing hierarchy) and integration with \textit{Roman's} Calibration Reference Data System to simulate pixel-level effects such as hot pixels. It will also produce data in the format of actual \textit{Roman} data, ASDF, allowing simulated data to be run through the same pipelines as real data.

We will produce a much larger dataset of at least 10$^5$ images using \textsc{mejiro} to be used in a data challenge. The objective of the data challenge will be to train generative neural networks that will expand simulated datasets, or to perform substructure inference using forward models or neural networks. 

%-----------------------------------------------------------------------
\section{Conclusion}
\label{section: Conclusion}

We have described a pipeline for generating a dataset of simulated \textit{Roman} WFI images of galaxy-galaxy strong gravitational lenses from a simulated HLWAS. We simulate a population of galaxies across cosmic time and select those forming detectable strong lenses in the design reference mission HLWAS. To study \textit{Roman's} impact on dark matter substructure inference with strong lenses, we add Cold Dark Matter subhalos. We leverage simulation tools based on \textit{Roman's} anticipated design and performance to add \textit{Roman} detector effects and create realistic simulated exposures. Assuming a fiducial case of single 146 s exposures, \textit{Roman's} HLWAS is estimated to find 160,000 detectable strong lenses, of which around 500 will be amenable to detailed substructure characterization. With coaddition and including the yield from the HLTDS, we expect \textit{Roman} to provide a sample of $\mathcal{O}(10^3)$ characterizable systems, enabling tighter constraints on the microphysics of dark matter. Owing to its wide field of view, the PSF varies across the WFI focal plane. For population-level substructure inference, we find that this variation has a non-negligible impact on milli-arcsecond scale lensing due to subhalos. Mitigating this effect by accounting for the PSF at the focal plane position a particular system was observed will be especially important for \textit{Roman} given how many systems it is expected to serendipitously observe. We characterize the field-dependence of the theoretical mass of the least massive detectable subhalo across the WFI focal plane. Isolating the effect of the PSF, we find that this mass varies by 5\% and tracks closely with overall wavefront error. We make our code and simulation products available to enable the generation and use of \textit{Roman} strong lenses for preparatory science.

%-----------------------------------------------------------------------
\section{Acknowledgments}

We thank the anonymous referee for valuable comments and feedback on our work. This research was supported by the National Aeronautics and Space Administration (NASA) under grant number 80NSSC24K0095 issued by the Astrophysics Division of the Science Mission Directorate (SMD). B.W. acknowledges support from NASA under a Future Investigators in NASA Earth and Space Science and Technology (FINESST) research grant (Grant No. 80NSSC24K1481). T.D. acknowledges support from the McDonnell Center for the Space Sciences at Washington University in St. Louis. X.H. acknowledges the USF Faculty Development Fund. P.N. gratefully acknowledges funding from the Department of Energy grant DE-SC0017660.

\software{\textsc{Astropy} \citep{Astropy2013, Astropy2018, Astropy2022}, \textsc{GalSim} \citep{Rowe2015}, \textsc{Lenstronomy} \citep{Birrer2018, Birrer2021}, \textsc{NumPy} \citep{Harris2020}, \textsc{phrosty}\footnote{\url{https://github.com/Roman-Supernova-PIT/phrosty}} (Aldoroty et al. 2025, in prep.), \textsc{pyHalo} \citep{Gilman2020}, \textsc{pysiaf} \citep{Osborne2024}, \textsc{SciPy} \citep{Virtanen2020}, \textsc{SkyPy} \citep{Amara2021, SkyPyCollaboration2023}, \textsc{SLSim} (DESC and SLSC Collaboration, in prep.), \textsc{speclite} \citep{Kirkby2024}, 
    \textsc{STIPS} \citep{Gomez2024},\footnote{\url{https://github.com/spacetelescope/STScI-STIPS/releases/tag/v2.1.0}} \textsc{WebbPSF} \citep{Perrin2012}.}

%-----------------------------------------------------------------------

\bibliography{references}{}
\bibliographystyle{aasjournal}

\end{document}